\documentclass[american]{elsarticle}
\usepackage[T1]{fontenc}
\usepackage[latin9]{inputenc}
\usepackage{textcomp}
\usepackage{graphicx}
\usepackage{amssymb}

\makeatletter
\newtheorem{theorem}{Theorem}
\newtheorem{definition}{Definition}

\makeatother

\usepackage{babel}

\begin{document}

\title{Exotic smooth $\mathbb{R}^{4}$, geometry of string backgrounds and
quantum D-branes}

\author{Torsten Asselmeyer-Maluga}

\address{German Aero space center, Rutherfordstr. 2, 12489 Berlin }

\ead{torsten.asselmeyer-maluga@dlr.de}

\author{Jerzy Kr\'ol}

\address{University of Silesia, Institute of Physics, ul. Uniwesytecka 4,
40-007 Katowice}

\ead{iriking@wp.pl}

\begin{abstract}
In this paper we make a first step toward determining 4-dimensional
data from higher dimensional superstring theory and considering these
as underlying structures for the theory. First, we explore connections
of exotic smoothings of $\mathbb{R}^{4}$ and certain configurations
of NS and D-branes, both classical and (generalized) quantum using
$C^{\star}$ algebras. Effects of some small exotic $\mathbb{R}^{4}$'s,
when localized on $S^{3}\subset\mathbb{R}^{4}$, correspond to stringy
geometries of $B$-fields on $S^{3}$. Exotic smoothness of $\mathbb{R}^{4}$
acts as a non-vanishing B-field on $S^{3}$ in $\mathbb{R}^{4}$.
The dynamics of D-branes in $SU(2)$ WZW model at finite $k$ indicates
the exoticness of ambient $\mathbb{R}^{4}$. 

Next, based on the relation of exotic smooth $\mathbb{R}^{4}$ with
integral levels of $SU(2)$ WZW model we show the correspondence between
exotic smoothness on 4-space, transversal to the world volume of NS5
branes, and the number of these NS5 branes. Relation with the calculations
in holographically dual 6-dimensional little string theory is discussed.

Generalized quantum D-branes in the noncommutative $C^{\star}$ algebras
corresponding to the codimension-1 foliations of $S^{3}$ are considered
and these determine the KK invariants of exotic smooth $\mathbb{R}_{h}^{4}$
for the case of non-integral $[h]\in H^{3}(S^{3},\mathbb{R})$. Moreover,
exotic smooth $\mathbb{R}^{4}$'s embedded in some exotic $\mathbb{R}^{4}$
as open submanifolds, are shown to correspond to generalized quantum
D-branes in the noncommutative $C^{\star}$ algebra of the foliation.
Finally, we show how exotic smoothness of $\mathbb{R}^{4}$ is correlated
with D6 brane charges in IIA string theory. 

In the last section we construct wild embeddings of spheres and relate
them to D-brane charges as well to KK theory. Wild embeddings as constructed
by using gropes are basic to understand exotic smoothness as well
Casson handles. Finally we conjecture that a quantum D-brane is wild
embedding. Then we construct an action for a quantum D-brane and show
that the classical limit (the usual embedding) agrees with the Born-Infeld
action.
\end{abstract}
\maketitle
\tableofcontents{}

\section{Introduction}

Despite the substantial effort toward quantizing gravity in 4 dimensions,
this issue is still open. One of the best candidates till now is the
superstring theory formulated in 10 dimensions. A way from superstring
theory to 4-dimensional quantum gravity or standard model of particle
physics (minimal supersymmetric extension thereof) is, at best, highly
non-unique. Many techniques of compactifications and flux stabilization
along with specific model-building branes configurations and dualities,
were worked out toward this end within the years. Possibly some important
data of a fundamental character are still missing. 

The point of view advocated in this paper is that indeed we have not
respected till now 4-dimensional phenomena of different smoothings
of Euclidean $\mathbb{R}^{4}$ which presumably are very important
for the program of QG. There are strong evidences that exotic 4-smoothness
on compact manifolds should be taken into account by any QG theory
\cite{Asselmeyer-Maluga2010}. Here we refer to open 4-manifolds and
try to consider exotic $\mathbb{R}^{4}$'s as serving a link between
higher dimensional superstring theory and 4-dimensional ,,physical\textasciiacute{}\textasciiacute{}
theories and 4-dimensional QG. String theory would describe directly
4-dimensional structures at the fundamental level. This paper serves
as a step toward seeing exotic smooth $\mathbb{R}^{4}$ as fundamental
objects underlying higher dimensional (super)string theory. Further
results regarding compactification and realistic 4-dimensional models
of various brane configurations in string theory and their relation
to exotic 4-smoothings, will be presented separately. 

The problem with successful inclusion of the effects of 4-open-exotics
into any physical theory, is the notorious lack of an explicit coordinate-like
description of these smooth manifolds. In the series of our recent
papers we addressed this issue and worked out some relative techniques
allowing for analytical treatment of small exotic $\mathbb{R}^{4}$'s
\cite{AsselmeyerKrol2009,AsselmeyerKrol2009a,AsselmeyerKrol2010,Krol2010}.
In this paper we show that description of D-branes in some exact string
backgrounds are related with 4-smoothness of $\mathbb{R}^{4}$. Moreover,
the deep quantum regime of the D-branes is also 4-exotic sensitive. 

However the connection of abstract, generalized quantum D-branes to
the actual superstring theory D-branes (in the manifold limit) is
not directly given. The Witten limit of superstring theory where D-branes
yield their noncommutative world-volumes is only the midway and in
fact motivates the full $C^{\star}$ algebra approach \cite{Szabo2008a}.
This last serves as a possible partial solution to the problem of
describing quantum D-branes in superstring theory. The connection
with exotic $\mathbb{R}^{4}$ at this, quantum level is unexpected
and shows that 4-dimensionality may get into the game in string theory
through back-door of nonperturbative quantum regime. In the last section
of the paper we use the $C^{*}$algebra approach to quantum D-branes
to construct a manifold model of a quantum D-brane as wild embedding.
Then we show that the $C^{*}$algebra of the wild embedding is isomorphic
to the $C^{*}$algebra of the quantum D-brane. Furthermore we construct
a quantum version of an action using cyclic cohomology and get the
right limit to the classical D-brane described by the Born-Infeld
action.

The basic technical ingredient of the analysis of small exotic $\mathbb{R}^{4}$'s
enabling uncovering many applications also in string theory is the
relation between exotic (small) $\mathbb{R}^{4}$'s and non-cobordant
codimension-1 foliations of the $S^{3}$ as well gropes and wild embeddings
as shown in \cite{AsselmeyerKrol2009}. The foliation are classified
by Godbillon-Vey class as element of the cohomology group $H^{3}(S^{3},\mathbb{R})$.
By using the $S^{1}$-gerbes it was possible to interpret the integral
elements $H^{3}(S^{3},\mathbb{Z})$ as characteristic classes of a
$S^{1}$-gerbe over $S^{3}$ \cite{AsselmeyerKrol2009a}. 

The main line of the topological argumentation can be briefly described
as follows:

\begin{enumerate}
\item In Bizacas exotic $\mathbb{R}^{4}$ one starts with the neighborhood
$N(A)$ of the Akbulut cork $A$ in the K3 surface $M$. The exotic
$\mathbb{R}^{4}$ is the interior of $N(A)$.
\item This neighborhood $N(A)$ decomposes into $A$ and a Casson handle
representing the non-trivial involution of the cork.
\item From the Casson handle we construct a grope containing Alexanders
horned sphere.
\item Akbuluts construction gives a non-trivial involution, i.e. the double
of that construction is the identity map.
\item From the grope we get a polygon in the hyperbolic space $\mathbb{H}^{2}$.
\item This polygon defines a codimension-1 foliation of the 3-sphere inside
of the exotic $\mathbb{R}^{4}$ with an wildly embedded 2-sphere,
Alexanders horned sphere.
\item Finally we get a relation between codimension-1 foliations of the
3-sphere and exotic $\mathbb{R}^{4}$.
\end{enumerate}
This relation is very strict, i.e. if we change the Casson handle
then we must change the polygon. But that changes the foliation and
vice verse. Finally we obtained the result:\\
\emph{The exotic $\mathbb{R}^{4}$ (of Bizaca) is determined by the
codimension-1 foliations with non-vanishing Godbillon-Vey class in
$H^{3}(S^{3},\mathbb{R}^{3})$ of a 3-sphere seen as submanifold $S^{3}\subset\mathbb{R}^{4}$.}

\section{Geometry of string backgrounds and exotic $\mathbb{R}^{4}$}

In this section we take the point of view that exotic smoothness of
some small exotic $\mathbb{R}^{4}$'s when localized on $S^{3}\subset\mathbb{R}^{4}$,
correspond to some stringy geometry given by so-called $B$-fields
on $S^{3}$. The localization is understood as the representation
of the exotics by 3-rd integral or real cohomologies of $S^{3}$.
This correspondence takes place in fact for a classical limit of the
geometry of string backgrounds, i.e. curved Riemannian manifold with
B-field. One can say that localized exotic smooth $\mathbb{R}^{4}$
on $S^{3}$ is described by stringy geometry of $B$-fields on this
$S^{3}$. The correspondence can be extended over string regime of
finite volume of $SU(2)$ WZW model.

\subsection{$SU(2)$ WZW model, D-branes and exotic $\mathbb{R}^{4}$}

We want to focus on changing the smoothness of $\mathbb{R}^{4}$ and
considering the changes as localized on $S^{3}$. As follows from
\cite{AsselmeyerKrol2009,AsselmeyerKrol2009a} this gives rise to
stringy effects, since the changes can be described by computations
in some 2D CFT, namely WZW models on $SU(2)$ at finite level. 

First we are going to discuss bosonic $SU(2)$ WZW model and dynamics
of branes in it. We deal here with $S^{3}$ hence the nonzero metric
of string background. In general, non-vanishing curvature $R(g)$,
where $g$ is a non-constant metric, of the background manifold $(M,g)$
on which bosonic string theory is formulated, enforces that $H$ -field
on $M$ cannot vanish. This is since the string field equations gives
rise to (see e.g. \cite{Schomerus2002}) \begin{equation}
R_{\mu\nu}(g)-\frac{1}{4}H_{\mu\rho\sigma}{H_{\nu}}^{\rho\sigma}={\cal O(\alpha')}\label{eq:R-H}\end{equation}
where $H=dB$ is the NSNS 3-form, $B=B(x)dx^{\mu}\wedge dx^{\nu}$
is the B-field, and dilaton is fixed to be constant. Also in the case
of superstring theory this equation still holds true provided all
RR background fields vanish \cite{Schomerus2002}.

D-branes in group manifold $SU(2)$ (at the semi-classical limit)
are determined as wrapping the conjugacy classes of $SU(2)$, which
are 2-spheres $S^{2}$ plus 2 points-poles, seen as degenerated 2-spheres.
Due to the quantization conditions there are $k+1$ D-branes on the
level $k$ $SU(2)$ WZW model \cite{Schomerus2000,Schomerus2002,Alekseev1999}.
To grasp the dynamics of the branes one should deal with the gauge
theory on the stack of $N$ D-branes on $S^{3}$ which is quite similar
to the flat space case where noncommutative gauge theory emerges \cite{Alekseev1999b}. 

For $N$ branes of type $J$ on top of each other, where $J$ is the
representation of $SU(2)_{k}$ i.e. $J=0,\frac{1}{2},1,\,...\,,\frac{k}{2}$,
the dynamics of the branes is described by the noncommutative action:

\begin{equation}
S_{N,J}=S_{YM}+S_{CS}=\frac{\pi^{2}}{k^{2}(2J+1)N}\left(\frac{1}{4}{\rm tr}(F_{\mu\nu}F^{\mu\nu})-\frac{i}{2}{\rm tr}(f^{\mu\nu\rho}{\rm CS}_{\mu\nu\rho})\right)\:.\label{eq:NoncommAction}\end{equation}
Here the curvature form $F_{\mu\nu}(A)=iL_{\mu}A_{\nu}-iL_{\nu}A_{\mu}+i[A_{\mu},A_{\nu}]+f_{\mu\nu\rho}A^{\rho}$
and the noncommutative Chern-Simons action reads ${\rm CS}_{\mu\nu\rho}(A)=L_{\mu}A_{\nu}A_{\rho}+\frac{1}{3}A_{\mu}[A_{\nu},A_{\rho}]$.
The fields $A_{\mu},\,\mu=1,2,3$ are defined on fuzzy 2-sphere $S_{J}^{2}$
and should be considered as $N\times N$ matrix-valued, i.e. $A_{\mu}=\sum_{j,a}{\rm a}_{j,a}^{\mu}Y_{a}^{j}$
where $Y_{a}^{j}$ are fuzzy spherical harmonics and ${\rm a}_{j,a}^{\mu}$
are Chan-Paton matrix-valued coefficients. $L_{\mu}$ are generators
of the rotations on fuzzy 2-spheres and they act only on fuzzy spherical
harmonics \cite{Schomerus2002}. The noncommutative action $S_{YM}$
was derived from Connes spectral triples of the noncommutative geometry
and was aimed to describe Maxwell theory on fuzzy spheres \cite{Watamura2000}. 

One can solve the equations of motion derived from the stationery
points of (\ref{eq:NoncommAction}) and the solutions describing the
dynamics of the branes, i.e. the condensation processes on the brane
configuration $(N,J)$ which results in another configuration $(N',J')$.
Namely the equation of motion derived from (\ref{eq:NoncommAction})
read:

\begin{equation}
L_{\mu}F^{\mu\nu}+[A_{\mu},F^{\mu\nu}]=0\label{eq:eof}\end{equation}
A class of solutions for (\ref{eq:eof}), in the semi-classical $k\to\infty$
limit, can be obtained from the $N(2J+1)$ dimensional representations
of the algebra ${\rm su}(2)$. For $J=0$ one has $N$ branes of type
$J=0$, i.e. $N$ point-like branes in $S^{3}$ at the identity of
the group. Given another solution corresponding to the $J_{N}=\frac{N-1}{2}$
one can show that this corresponds to the brane wrapping the $S_{J_{N}}^{2}$
sphere and is obtained as the condensed state of $N$ point-like branes
at the identity of $SU(2)$ \cite{Schomerus2002}:

\begin{equation}
(N,J)=(N,0)\to(1,\frac{N-1}{2})=(N',J')\label{eq:semi-class:N-to-N1}\end{equation}

Turning to the finite $k$ stringy regime of the $SU(2)$ WZW model
one can make use of the techniques of the boundary CFT when applied
to the analysis of Kondo effect \cite{Schomerus2002}. It follows
that there exists a continuous shift at the level of partition function,
between $N\chi_{j}(q)$ and the interfered sum of characters $\sum_{j}N_{J_{N}j}^{\; l}\chi_{l}(q)$
where $N=2J_{N}+1$ (in the vanishing value of the coupling constant)
and $N_{J_{N}j}^{\; l}$ are Verlinde fusion rule coefficients. In
the case of $N$ point-like branes one can determine the decay product
of these by considering open strings ending on the branes. The result
on the partition function is

\[
Z_{(N,0)}(q)=N^{2}\chi_{0}(q)\]
which is continuously shifted to $N\chi_{J_{N}}(q)$ and next to $\sum_{j}N_{J_{N}J_{N}}^{\;\; j}\chi_{j}(q)$.
As the result we have the decay process \cite{Schomerus2002}

\begin{equation}\label{eq:stringN-to-N1}
\begin{array}{c}
Z_{(N,0)}(q)\to Z_{(1,J_{N})}\\[4pt]

(N,0)\to(1,J_{N})
\end{array}
\end{equation}which extends the similar process derived at the semi-classical $k\to\infty$
limit in the effective gauge theory (\ref{eq:semi-class:N-to-N1}),
however the representations $2J_{N}$ are bounded now, from the above,
by $k$.

Given the above dynamics of branes in the WZW $SU(2)$ model at stringy
regime, one can address the question of brane charges in a direct
way. This is based on the decay rule (\ref{eq:stringN-to-N1}) in
the supersymmetric WZW $SU(2)$ model. In this case we have a shift
of the level namely $k\to k+2$ which measures the units of the NSNS
flux through $SU(2)=S^{3}$. One can see the supersymmetric model
as strings moving on $SU(2)$ with $k+2$ units of NSNS flux. From
the CFT point of view there exist currents $J^{a}$ which satisfy
$k+2$ level of the Kac-Moody algebra and free fermionic fields $\psi^{a}$
in the adjoint representation of $su(2)$. However it is possible
to redefine the bosonic currents as

\[
J^{a}+\frac{i}{k}f_{\: bc}^{a}\psi^{b}\psi^{c}\]
which fulfill the current algebra commutation relation at the level
$k$. Here $f_{\: bc}^{a}$ are the structure constants of $su(2)$.
The fields $\psi^{a}$ commute with such currents, thus we have the
splitting of the supersymmetric WZW $SU(2)$ model at level $k+2$
as WZW $SU(2)$ model at level $k$ times the theory of free fermionic
fields. 

Thus there are $k+1$ stable branes wrapping the conjugacy classes
numbered by $J=0,\frac{1}{2},...,\frac{k}{2}$. The decaying process
(\ref{eq:stringN-to-N1}) says that placing $N$ point-like branes
(each charged by the unit $1$) at the pole $e$ they can decay to
the spherical brane $J_{N}$ wrapping the conjugacy class. Taking
more point-like branes to the stack at $e$ gives the more distant
$S^{2}$ branes until reaching the opposite pole $-e$ where we have
single point-like brane with the opposite charge $-1$. Having identify
$k+1$ units of the charge with $-1$ we arrive at the conclusion
that the group of charges is $\mathbb{Z}_{k+2}$. More generally the
charges of branes on the background $X$ with non-vanishing $H\in H^{3}(X,\mathbb{Z})$
are described by the twisted $K$ group $K_{H}^{\star}(X)$ (see e.g.
\cite{MathaiMurray2001}). In the case of $SU(2)$ we get the group
of RR charges as above for $K=k+2$ \begin{equation}
K_{H}^{\star}(S^{3})=\mathbb{Z}_{K}\end{equation}

Based on \cite{AsselmeyerKrol2009}, the following important observation
is in order:\emph{ certain small exotic $\mathbb{R}^{4}$'s generate
the group of RR charges of D-branes in the curved background of $S^{3}\subset\mathbb{R}^{4}$.}
This observation is based on the integral classes $H\in H^{3}(S^{3},\mathbb{Z})$
from which one can construct the exotic $\mathbb{R}_{H}^{4}$ as corresponding
to the codimension-1 foliation of $S^{3}$ (determined by the class
$H$). In \cite{AsselmeyerKrol2009} we showed that twisted K-theory
of $S^{3}$ by the class $H\in H^{3}(S^{3},\mathbb{Z})$ can be seen
as the effect of the exotic smoothness $\mathbb{R}_{H}^{4}$ on the
ambient 4-space, when $S^{3}$ is understood as the part of the boundary
of the Akbulut cork of $\mathbb{R}_{H}^{4}$.

Thus we arrive at the correspondence:

\begin{theorem}

The classification of RR charges of the branes on group manifold background
$SU(2)$ at the level $k$, hence the dynamics of D-branes in $S^{3}$
in stringy regime, is correlated with exotic smoothness on $\mathbb{R}^{4}$
containing this $S^{3}=SU(2)$ as the part of the boundary of the
Akbulut cork. 

\end{theorem} 

We can give yet another interpretation of the 4-exoticness which appears
on flat $\mathbb{R}^{4}$ in this context. Exotic smoothness of $\mathbb{R}^{4}$,
$\mathbb{R}_{H}^{4}$, determines the collection of stable D-branes
on $SU(2)$ at the level $k$ of the WZW model, where $k=[H]\in H^{3}(S^{3},\mathbb{Z})$.
Thus, the stringy, finite $k$, level of WZW model characterizes exotic
4-smoothness. Recall that in the case of $H=0$ (e.g. $B$ constant
in a flat space, i.e. in $k\to\infty$ limit) the smooth structure
on $\mathbb{R}^{4}$ is the standard one \cite{AsselmeyerKrol2009}.
Thus the exotic smoothness on $\mathbb{R}^{4}$ translates the 4-curvature
to the non-zero H-field on $S^{3}$ of finite volume in string units.
This is similar to the effect of string field equations relating $R$
and $H$ as in (\ref{eq:R-H}), though it holds now between different
spaces ($\mathbb{R}^{4}$ and $S^{3}$).

\subsection{$SU(2)$ WZW model in the geometry of the stack of NS5-branes}

The group manifold $SU(2)=S^{3}$ is the only manifold which became
relevant so far for the description of small exotic $\mathbb{R}^{4}$.
From the other side it is the only one which appears directly as part
of a string background (namely one generated by NS5-branes). The reason
is given by the connection of 4-exotics and string theory as it can
be naturally formulated in the geometry of the stack of NS5-branes.
Let us briefly describe this string background \cite{Schomerus2000,Schomerus2002,Bachas2000}.

We consider a configuration of $k$ coincident supersymmetric NS5-branes
in type II theory. The full fivebrane background is (in string frame) 

\begin{equation}\label{eq:NS5-background}
\begin{array}{c}
ds^{2}=dx^{2}+f(r)dy^{2}\\[4pt]

e^{2\phi}=g_{s}^{2}f(r)\\[4pt]

f(r)=1+\frac{k\alpha'}{r^{2}} \\[4pt]

H_{IJK}=k\alpha'\epsilon_{IJK}
\end{array}
\end{equation}where $x$ are the $5+1$ longitudinal coordinates along NS5-branes
referred to by indices $\mu$, $\nu$, etc., $y$ being 4 transverse
coordinates referred to by indices $I$, $J$, $K$ ... and $r=|y|$,
$1/\alpha'\sim$ string tension. The fields of this background reads
as

\begin{equation}
\begin{array}{c}
e^{2\Phi}=1+\sum_{j=1}^{k}\frac{l_{s}^{2}}{|y-y_{j}|^2}\\[4pt]

g_{IJ}=e^{2\Phi}\delta_{IJ}\\[4pt]

g_{\mu\nu}=\eta_{\mu\nu} \\[4pt]

H_{IJK}=-\epsilon_{IJKL}\partial^{L}\Phi
\end{array}
\end{equation}where $y_{j},j=1,...,k$ are the positions of the NS5-branes. When
the branes coincide at 0, $y_{j}=0$, the near horizon solutions $y\to0$,
are

\begin{equation}
\begin{array}{c}
e^{2\Phi}=\frac{kl_{s}^{2}}{|y|^{2}}\\[4pt]

g_{IJ}=e^{2\Phi}\delta_{IJ}\\[4pt]

g_{\mu\nu}=\eta_{\mu\nu} \\[4pt]

H_{IJK}=-\epsilon_{IJKL}\partial^{L}\Phi
\end{array}
\end{equation}

In the near-horizon limit $r=|y|^{2}\to0$, the background factorizes
into a radial component and a $S^{3}$ and flat 6-dimensional Minkowski
spacetime. Strings propagating at this limiting background are described
by the exact world-sheet CFT with the target $\mathbb{R}^{5,1}\times\mathbb{R}_{\phi}\times S_{k}^{3}$.
Here $\mathbb{R}_{\phi}$ is the real line with the parameter $\phi$
which is a scalar corresponding to the ,,linear dilaton''

\begin{equation}
\begin{array}{c}
\Phi=-\sqrt{\frac{1}{2k}}\phi\\[4pt]

\phi=\sqrt{\frac{k}{2}}\log\frac{r}{kl_{s}^{2}}

\end{array}
\end{equation}

The flat Minkowski space $\mathbb{R}^{5,1}$ is longitudinal to the
directions of NS5-branes, $S_{k}^{3}$ is $SU(2)_{k}$ and is a level
$k$ WZW supersymmetric CFT (SCFT) on $SU(2)$ as discussed in the
previous subsection. This $S^{3}$ corresponds to the angular coordinates
of the transversal $\mathbb{R}^{4}$. We see that infinite geometrical
,,throat'' $\mathbb{R}_{\phi}\times S_{k}^{3}$, emerges. The metric
of the background (in the string frame) thus reads

\[
ds^{2}=dx_{6}^{2}+d\phi^{2}+kl_{s}d\Omega_{3}^{2}\,,\: g_{s}^{2}(\phi)=e^{-2\phi/\sqrt{k}l_{s}}\:.\]
This background is obtained in the near horizon, $\phi\to-\infty$
($r\to0$), geometry of the stack of $k-2$ NS5-branes in type II
string theory and is in fact a SCFT on the throat. The NS5-branes
are placed at $\phi\to-\infty$ and string theory is strongly coupled
there, $g_{s}\sim\exp(2\Phi)$. In the opposite limit $\phi\to+\infty$,
or $r\to+\infty$, gives asymptotically flat 10-space and string theory
is weakly coupled in that limit. This is essentially the CHS (Callan,
Harvey, Strominger \cite{CHS1991}) exact string theory background
where $SU(2)$ WZW model appears at suitable level $k$. 

Given the CHS limiting geometry of $N$ NS5-branes we have the 4-dimensional
tube $\mathbb{R}_{\phi}\times S^{3}$. The volume of $S^{3}$ in string
units is finite and correlated with the number of NS5-branes by $N=k-2$
\cite{Bachas2000}. We take an exotic $\mathbb{R}_{H}^{4}$ for $[H]=k\in H^{3}(S^{3},\mathbb{Z})$.
This can be achieved more directly by considering the Akbulut cork
$A_{H}$ with the boundary, $\partial A_{H}=\Sigma_{H}$, the homology
3-sphere. As was shown in \cite{AsselmeyerKrol2009} $\Sigma_{H}$
contains $S^{3}$ such that the codimension-1 foliations of it generates
the foliations of $\Sigma_{H}$. The foliations in turn are generated
by Casson handles attached to $A$. Thus the attached Akbulut cork
and Casson handle(s) determine the small exotic smoothness of $\mathbb{R}_{H}^{4}$
\cite{GomSti:1999,AsselmeyerKrol2009}. Moreover, the cobordism classes
of codimension-1 foliations of $S^{3}$ are classified by the Godbillon-Vey
invariants which are elements of $H^{3}(S^{3},\mathbb{R})$. In our
case we deal with integral 3-rd cohomologies $[H]\in H^{3}(S^{3},\mathbb{Z})$.
Thus, a way of embedding the Akbulut cork, for some class of exotic
$\mathbb{R}^{4}$'s, in the ambient $\mathbb{R}^{4}$ is determined
by the integral classes $k\in H^{3}(S^{3},\mathbb{Z})$. Taking the
above $S^{3}$ from the boundary of the Akbulut cork, as $S^{3}=SU(2)$
in the string background of $N$ NS5-branes we arrive at the following
result:

\begin{theorem}\label{Th:Ns5Branes}

In the geometry of the stack of NS5-branes in type II superstring
theories, adding or subtracting a NS5-brane is correlated with the
change of smoothing on transversal $\mathbb{R}^{4}$. 

\end{theorem}

Now the tube $\mathbb{R}_{\phi}\times S_{k}^{3}$ of the limiting
geometry can be embedded in the ambient standard $\mathbb{R}^{4}$.
Taking this $S_{k}^{3}$ as lying in the boundary of the Akbulut cork
for some exotic smooth $\mathbb{R}_{H}^{4}$, the embedding of the
tube in this exotic 4-space is determined by the embedding of the
Akbulut cork. But this embedding is determined by Casson handles attached
to the cork and corresponds to the integral class $[H]=k\in H^{3}(S^{3},\mathbb{Z})$.
Thus the background $\mathbb{R}^{5,1}\times\mathbb{R}_{\phi}\times SU(2)_{k}$
is geometrically realized as $\mathbb{R}^{5,1}\times\mathbb{R}_{H}^{4}$.
We propose here a general heuristic rule:

R1. \emph{D-branes probing exotic 4-dimensional Euclidean space, $\mathbb{R}_{H}^{4}$,
times 6-dim. Minkowski spacetime ,$M^{5,1}$, are described equivalently
by the D-branes of type II string theory probing the transversal 4-space,
$\mathbb{R}^{4}$, to $k$ NS5-branes in the background of these 5-branes.
Here $[H]=k\in H^{3}(S^{3},\mathbb{Z})$. Since $M^{5,1}$ appears
in both sides of the correspondence we say that }D-branes explore
exotic Euclidean $\mathbb{R}_{H}^{4}$\emph{. }

Rule R1 is based on the assumption that various nonstandard smoothings
of $\mathbb{R}^{4}$ can be grasped by the effects of $H^{3}(S^{3},\mathbb{Z})$.
This follows from the correlation of the classes and 4-exotics as
proved in \cite{AsselmeyerKrol2009}. Following this rule we can consider
many examples of D-branes in the above background (see e.g. \cite{GiveonAntoniadis2000,GiveonKutasov2000,YunKwon2009,Ribault2003,ChenSun2005}),
as referring to 4-exoticness. 

Furthermore type II string theory on $\mathbb{R}^{5,1}\times\mathbb{R}_{\phi}\times SU(2)_{k}$
is given by the SCFT on the infinite ,,throat'' of the background,
i.e. $\mathbb{R}_{\phi}\times S^{3}$. Then this theory was proposed
to be approachable via holography by using duality. The holographically
dual theory appears to be so called 6-dimensional \emph{little string
theory (LST)} \cite{GiveonKutasov2000,Aharony2002}. This is a very
interesting situation for us since LST was analyzed as having possible
experimental signatures at the TeV scales after the compactification
on torus \cite{GiveonAntoniadis2000}. By the rule above this refers
to 4-exotics as well. We do not deal here with the details and refer
the interesting reader to a separate paper devoted to (flux) compactification
in string theory and exotic 4-smoothness. But we will present some
general remarks here.

LST's are non-local theories without gravity and can be described
in the limit $g_{s}\to0$ in the theory on $k$ NS5-branes. In that
limit the bulk degrees of freedom decouple, hence gravity does. This
6-dim. LST without gravity is holographically dual to the type II
string theory formulated on the background $\mathbb{R}^{5,1}\times\mathbb{R}_{\phi}\times SU(2)_{k}$
\cite{GiveonAntoniadis2000}. From the rule R1 it follows that LST
is referred to exotic $\mathbb{R}_{H}^{4}$ and calculations in LST
should lead to invariants of the 4-exotics. The perturbative calculations,
however, are hardly performed in LST since the string coupling $g_{s}$
diverges in the dual string background along the tube, and LST is
sensitive on that. One usually regulates the geometry via chopping
the tube. But the decomposition of the SCFT $SU(2)_{k}$ on $S_{Y}^{1}\times SU(2)_{k}/U(1)$
can be performed. Here $SU(2)_{k}/U(1)$ is a minimal $N=2$ model
at the level $k$ and $S_{Y}^{1}$ is the Cartan subalgebra of $SU(2)$
with the parameter $Y$. The dependence on $k$ is crucial at this
reformulation since this refers to 4-exotics by theorem \ref{Th:Ns5Branes}
and the rule R1. Thus we have the SCFT $\mathbb{R}_{\phi}\times S_{Y}^{1}\times\frac{SU(2)_{k}}{U(1)}$
instead of the tube $\mathbb{R}_{\phi}\times SU(2)_{k}$. The chopping
of the strong coupling region is now performed by taking the SCFT
$\frac{SL(2)_{k}}{U(1)}$ instead of $\mathbb{R}_{\phi}\times S_{Y}^{1}$
which means replacing the background $\mathbb{R}^{5,1}\times\mathbb{R}_{\phi}\times SU(2)_{k}$
by $\mathbb{R}^{5,1}\times\frac{SL(2)_{k}}{U(1)}\times\frac{SU(2)_{k}}{U(1)}$.
This means, on the level of $k$ NS5-branes, the separation of these
5-branes along the transverse circle of radius $L$. Now the double-scale
limit of LST is the one when taking both $g_{s}$ and $L$ to zero
while $\frac{L}{g_{s}}$ remains constant. 

Following \cite{GiveonKutasov2000} we can take systems of D4, D6-branes
between separated NS5-branes. The various expressions like correlation
functions can be now calculated perturbatively in the holographically
dual 6-dimensional LST theory. Besides, suitable compactifications
may refer to the spectra with the TeV scale of the standard model
of particles. The dependence on $k$ of some of these expressions
can be seen as the signature of the existence of exotic structure
in the 4-space transversal to the branes.

\emph{Exoticness of the 4-space transversal to the worldvolume of
NS5-branes, is reflected in specific perturbative spectra of D-branes
when calculated in dual 6-dimensional LST. When compactifying this
LST on 2 directions longitudinal to the 5-brane one gets spectra which
could be sensitive on transversal exoticness of $\mathbb{R}^{4}$.
From the point of view of physics, the calculations refer to the TeV
scale \cite{Aharony2002}. }

The important observation can be made: \emph{Some LST calculations
refer not only to holographically dual string theory but also to exotic
smoothness on $\mathbb{R}^{4}$}. \emph{This is the indication that
one can try, at least in some cases, to replace higher dimensional
string theory effects by 4-dimensional phenomena.} 

This is in fact the reformulation of the rule R1. The NS5-branes backgrounds
show that string theory computations ,,feel'' the 4-exoticness.

\section{Quantum D-branes and 4-exotica}

In this section we want to show that D-branes of string theory, as
in the previous sections, are related with exotic smooth $\mathbb{R}^{4}$'s
also beyond the semi-classical limit, i.e. in the quantum regime of
the theory where one should deal rather with \emph{quantum branes}.
What \emph{quantum branes} mean in general is still an open and hard
problem. One appealing proposition, relevant for this paper, is to
consider branes in noncommutative spacetimes rather than on commutative
manifolds or orbifolds. This leads to abstract D-branes in general
noncommutative separable $C^{\star}$ algebras as counterparts for
quantum D-branes. In the next section we will present a definition
using wild embeddings.

\subsection{D-branes on spaces: K-homology and KK theory \label{sub:D-branes-on-spaces:} }

The description of systems of stable Dp-branes of IIA,B string theories
via K-theory of topological spaces can be extended toward the branes
in noncommutative $C^{\star}$ algebras. A direct string representation
of the algebraic and K-theoretic ideas is best seen in K-matrix string
theory where, in particular, tachyons are elements of the spectral
triples representing the noncommutative geometry of the world-volumes
of the configurations of branes \cite{AsakawaSugimotoTerasima2002}.
The elements of the formulation of type II strings as K matrix theory
is presented in the Appendix \ref{sec:Elements-of-K-matrix}.

First let us consider the case of vanishing $H$-field on $X$. The
charges of D-branes are classified by suitable $K$ theory groups,
i.e. $K^{0}(X)$ in IIB and $K^{1}(X)$ in IIA string theories, where
$X$ is the background manifold.  On the other hand, world-volumes
of Dp-branes correspond to the cycles of K homology groups, $K_{1}(X)$,
$K_{0}(X)$, which are dual to the $K$ theory groups. Let us see
how $K$ -cycles correspond to the configurations of D-branes. 

A $K$ - cycle on $X$ is a triple $(M,E,\phi)$ where $M$ is a compact
${\rm {Spin}^{c}}$ manifold without boundary, $E$ is a complex vector
bundle on $M$ and $\phi:M\to X$ is a continuous map. The topological
$K$-homology $K_{\star}(X)$ is the set of equivalence classes of
the triples $(M,E,\phi)$ respecting the following conditions:

\begin{itemize}

\item[(i)] $(M_{1},E_{1},\phi_{1})\sim(M_{2},E_{2},\phi_{2})$ when
there exists a triple (bordism of the triples) $(M,E,\phi)$ such
that $(\partial M,E_{|\partial M},\phi_{|\partial M})$ is isomorphic
to the disjoint union $(M_{1},E_{1},\phi_{1})\cup(-M_{2},E_{2},\phi_{2})$
where $-M_{2}$ is the reversed ${\rm {Spin}^{c}}$ structure of $M_{2}$
and $M$ is a compact ${\rm {Spin}^{c}}$ manifold with boundary. 

\item[(ii)] $(M,E_{1}\oplus E_{2},\phi)\sim(M,E_{1},\phi)\cup(M,E_{2},\phi)$,

\item[(iii)] Vector bundle modification $(M,E,\phi)\sim(\widehat{M},\widehat{H}\otimes\rho^{\star}(E),\phi\circ\rho)$.
$\widehat{M}$ is even dimensional sphere bundle on $M$, $\rho:\widehat{M}\to M$
projection, $\widehat{H}$ is a vector bundle on $\widehat{M}$ which
gives the generator of $K(S_{q}^{2p})=\mathbb{Z}$ on every $S_{q}^{2p}$
over each $q\in M$ \cite{Szabo2002a}. 

\end{itemize}

The topological K-homology as above has an abelian group structure
with disjoint union of cycles as sum. The triples $(M,E,\phi)$ with
$M$ being even dimensional determines $K_{0}(X)$. Similarly, $K_{1}(X)$
corresponds to odd dimensions. Thus $K_{\star}(X)$ decomposes into
a direct sum of abelian groups:

\[
K_{\star}(X)=K_{0}(X)\oplus K_{1}(X)\,.\]

Now the interpretation of cycles $(M,E,\phi)$ as D-branes \cite{HarveyMoore2000}
is the following: $M$ is the world-volume of brane, $E$ the Chan-Paton
bundle on it and $\phi$ gives the embedding of the brane into spacetime
$X$. Moreover, $M$ has to wrap ${\rm Spin}^{c}$ manifold \cite{FreedWitten1999}
and $K_{0}(X)$ classifies stable D-branes configurations in IIB,
and $K_{1}(X)$ in IIA, string theories. The equivalences of K-cycles
as formulated in the conditions (i)-(iii) correspond to natural relations
for D-branes \cite{AsakawaSugimotoTerasima2002,Szabo2008b}. 

The topological K-homology theory above can be obtained analytically
(analytic K-homology theory) as a special commutative case of the
following construction on general $C^{\star}$ algebras \cite{AsakawaSugimotoTerasima2002}.

A Fredholm module over a $C^{\star}$ algebra ${\cal A}$ is a triple
$({\cal H},\phi,F)$ such that 

\begin{enumerate}
\item ${\cal H}$ is a separable Hilbert space,
\item $\phi$ is a $^{\star}$ homomorphism between $C^{\star}$ algebras
${\cal A}$ and ${\rm {\bf B}}({\cal H})$ of bounded linear operators
on ${\cal H}$,
\item $F$ is self-adjoint operator in ${\rm {\bf B}}({\cal H})$ satisfying
\end{enumerate}
\[
F^{2}-id\in{\rm K}({\cal H})\,,\quad[F,\phi(a)]\in{\rm K}({\cal H})\:{\rm for}\:{\rm every}\: a\in{\cal A}\]
where ${\rm K}({\cal H})$ are compact operators on ${\cal H}$. Now
let us see how a Fredholm module $({\cal H},\phi,F)$ describes certain
configuration of IIA K matrix string theory directly related to D
branes. To this end we consider the operators of the K-matrix theory
$\Phi^{0},...,\Phi^{9}$ (infinite matrices) acting on the Hilbert
space ${\cal H}$ as generating the $C^{\star}$ algebra ${\cal A}_{M}$
(see the Appendix \ref{sec:Elements-of-K-matrix} and \cite{AsakawaSugimotoTerasima2002}).
In the case of commuting $\Phi^{\mu}$, hence commutative ${\cal A}_{M}$,
we have the following correspondence (explaining the index $M$ in
${\cal A}_{M}$): 

\begin{itemize}
\item Every commutative $C^{\star}$ algebra is isomorphic to the algebra
of continuous complex functions vanishing at infinity $C(M)$ on some
locally compact Hausdorff space $M$ (Gelfand-Najmark theorem). A
point $x\in M$ is determined by a character of ${\cal A}_{M}$ which
is a $^{\star}$ homomorphism $\phi_{x}:{\cal A}_{M}\to\mathbb{C}$.
\item $M$ serves as a common spectrum for $\Phi^{0},...,\Phi^{9}$ and
the choice of a point from $M$ as the eigenvalue of $\Phi^{\mu}$
fixes the position of the non BPS instanton along $x^{\mu}$.
\item In this way $M$ is covered by the positions of infinite many non
BPS instantons and serves as the world-volume of some higher dimensional
D brane \cite{AsakawaSugimotoTerasima2002}.
\end{itemize}
Now let us explain the role of the tachyon $T$. $T$ is a self-adjoint
unbounded operator acting on the Chan-Paton Hilbert space ${\cal H}$.
${\cal A}_{M}$ is a $C^{\star}$ unital algebra generated by $\Phi^{0},...,\Phi^{9}$
which can be now noncommutative. The corresponding geometry of the
world-volume $M$ would be noncommutative and given by some spectral
triple. The spectral triple is in fact $({\cal H},{\cal A},T)$ which
means that the following conditions are satisfied \cite{AsakawaSugimotoTerasima2002}:

\[
T-\lambda\in{\rm {\bf K}}({\cal H})\:{\rm for}\:{\rm every}\:\lambda\in\mathbb{C}\setminus\mathbb{R},\;[a,T]\in{\bf B}({\cal H})\:{\rm for}\:{\rm every}\: a\in{\cal A}_{M}\]
These conditions indeed hold true in our case of K matrix string theory
for a tachyon field $T$, Chan-Paton Hilbert space ${\cal H}$ and
$C^{\star}$ algebra ${\cal A}_{M}$ generated by $\Phi^{\mu}$ (see
Appendix \ref{sec:Spectral-triples-and}). The extension of spacetime
manifold toward noncommutative algebra and noncommutative world-volumes
of branes, represented by spectral triples, is thus given by \cite{AsakawaSugimotoTerasima2002}:

\begin{enumerate}
\item Fixing the spacetime $C^{\star}$ algebra ${\cal A}$;
\item A $^{\star}$ homomorphism $\phi:{\cal A}\to{\bf B}({\cal H})$ generates
embedding of the D-brane world-volume $M$ and its noncommutative
algebra ${\cal A}_{M}$ as ${\cal A}_{M}:=\phi({\cal A})$;
\item D-branes embedded in a spacetime ${\cal A}$ are represented by the
spectral triple $({\cal H},{\cal A}_{M},T)$;
\item Equivalently, D-brane in $A$ is given by unbounded Fredholm module
$({\cal H},\phi,T)$.
\end{enumerate}
In particular the classification of stable D-branes in ${\cal A}$
is the classification of Fredholm modules $({\cal H},\phi,T)$ given
by analytical K-homology. Given the isomorphisms of the topological
and analytical K homology groups, we have the classification of stable
D-branes in terms of K-cycles, as we discussed at the beginning of
this section. In terms of K matrix string theory we can say that stable
configurations of D-instantons determine the stable higher dimensional
D-branes which are K-homologically classified as above. 

Now let us turn to a more general situation than K-string theory of
D-instantons, i.e. backgrounds given by non-BPS Dp-branes or non-BPS
Dp-${\rm \overline{{\rm Dp}}}$-branes in type II string theory. The
stable configurations of Dq-branes are then classified by generalized
K-theory namely Kasparov KK-theory. As in the above case of D-branes
in a $C^{\star}$ algebra ${\cal A}$ corresponding to Fredholm modules,
one defines an odd Kasparov module $({\cal H}_{{\cal B}},\phi,T)$,
where ${\cal H}_{{\cal B}}$ is an countable Hilbert module over $C^{\star}$algebra
${\cal B}$, as

\begin{itemize}
\item a $\star$-homomorphism from ${\cal A}$ to the $C^{\star}$ algebra
of bounded linear operators on ${\cal H}_{{\cal B}}$, $\phi:{\cal A}\to{\rm {\bf B}}({\cal H}_{{\cal B}})$;
\item a self-adjoint operator $T$ from ${\rm {\bf B}}({\cal H}_{{\cal B}})$
satisfying: 
\end{itemize}
\[
T^{2}-1\in{\rm {\bf K}}({\cal H}_{{\cal B}})\:{\rm and}\:[T\,,\phi(a)]\in{\rm {\bf K}}({\cal H}_{{\cal B}})\:{\rm for}\,{\rm every}\, a\in{\cal A}\,,\]
where ${\rm {\bf K}}({\cal H}_{{\cal B}})$ is ${\cal B}\otimes{\bf {\rm K}}$.
$({\cal H}_{{\cal B}},\phi,T)$ is in fact a family of Fredholm modules
on the algebra ${\cal B}$. When ${\cal B}$ is $\mathbb{C}$ we have
an ordinary Fredholm module as before. The homotopy equivalence classes
of odd Kasparov modules $({\cal H}_{{\cal B}},\phi,T)$ determine
elements of $KK^{1}({\cal A},{\cal B})$. Also one defines an even
Kasparov classes $KK^{0}({\cal A},{\cal B})=KK({\cal A},{\cal B})$
as homotopy equivalence classes of the triples $({\cal H}_{{\cal B}}^{(0)}\oplus H_{{\cal B}}^{(1)},\phi^{(0)}\oplus\phi^{(1)},\left(\begin{array}{cc}
0 & T^{\star}\\
T & 0\end{array}\right))$. A natural $\mathbb{Z}_{2}$ grading appears due to the involution
${\cal H}_{{\cal B}}^{(0)}\oplus H_{{\cal B}}^{(1)}\to{\cal H}_{{\cal B}}^{(0)}\oplus-H_{{\cal B}}^{(1)}$. 

Now the classification pattern for branes in spaces emerges. There
are non-BPS unstable Dp-branes wrapping the $p+1$-dimensional world-volume
$B$. Then stable Dq-branes configurations embedded in a space $A$
transverse to $B$ correspond to (are classified by) the classes of
$KK^{1}(A,B)$. Similarly, given non-BPS unstable Dp-${\rm \overline{Dp}}$-branes
system, then stable Dq-branes embedded in $A$ transverse to $B$
($p+1$-dimensional world-volumes) are classified by elements of $KK^{0}(A,B)$.
The case of even $KK^{0}(A,B)$ contains the $\mathbb{Z}_{2}$ grading
as corresponding to the Chan-Paton indices of Dp and ${\rm \overline{Dp}}$-branes.

\subsection{D-branes on separable $C^{\star}$ algebras and KK theory \label{sub:Branes-on-separable}}

The classification of D-branes in a spacetime manifold given by KK
theory as sketched in the previous subsection, can be extended over
noncommutative spacetimes and noncommutative D-branes both represented
by separable $C^{\star}$ algebras. Let us first recapitulate the
,,classic'' case of spaces allowing the extension over $C^{\star}$
algebras \cite{Szabo2008c}. 

In the case of type II superstring theory, let $X$ be a compact part
of spacetime manifold, i.e. $X$ is a compact ${\rm spin}^{c}$ manifold
again with no background $H$-flux. As we saw, a flat D-brane in $X$
is a Baum-Douglas K-cycle $(W,E,f)$. Here $f:W\hookrightarrow X$
is the embedding of the closed ${\rm spin}^{c}$ submanifold $W$
of $X$ and $E\to W$ is a complex vector bundle with connection (Chan-Paton
gauge bundle). As follows from Baum-Douglas construction, $E$ determines
the stable class in the K-theory group $K^{0}(W)$ and all K-cycles
form an additive category under disjoint union. Now, the set of all
K-cycles classes up to a kind of gauge equivalence as in Baum-Douglass
construction, gives the K-homology of $X$. This K-homology is also
the set of stable homotopy classes of Fredholm modules which are taken
over the commutative $C^{\star}$ algebra $C(X)$ of continuous functions
on $X$. This defines the correspondence (isomorphism) where a K-cycle
$(W,E,f)$ corresponds to unbounded Fredholm module $({\cal H},\rho,D_{E}^{\mbox{W}})$.
Here ${\cal H}$ is the separable Hilbert space of square integrable
spinors on $W$ taking values in the bundle $E$, i.e. $L^{2}(W,S\otimes E)$,
$\rho:C(X)\to{\rm {\bf B}}({\cal H})$ is the representation of the
$C^{\star}$ algebra $C(X)$ in ${\cal H}$ such that $C(X)\ni g\to a_{g\circ f}\in{\rm {\bf B}}({\cal H})$
where $a_{g\circ f}$ is the operator of point-wise multiplication
of functions in $L^{2}(W,S\otimes E)$ by the function on $W$, $g\circ f$,
and $f:W\hookrightarrow X$. $D_{E}^{W}$ is the Dirac operator twisted
by $E$ corresponding to the ${\rm spin}^{c}$ structure on $W$.
Given the K-theory class of the Chan-Paton bundle $E$, i.e. $[E]\in K^{0}(W)$,
then the dual K-homology class of a D-brane, $[W,E,f]$ uniquely determines
$[E]$. In that way D-branes determine K-homology classes on $X$
which are dual to K-theory classes from $K^{r}(X)$ where $r$ is
the transversal dimension for the brane world-volume $W$. This K-theory
class is derived from the image of $[E]\in K^{0}(W)$ by the Gysin
K-theoretic map $f_{!}$. As we discussed already, the odd and even
classes of K-homology $K_{\star}(X)$ correspond to the parity of
the dimension of $W$. The K-cycle $(W,E,f)$ corresponds to a Dp-brane
and its gauge equivalence is given by Baum-Douglas construction using
the conditions (i)-(iii) in Sec. \ref{sub:D-branes-on-spaces:}. Thus
we have \cite{Szabo2008b}:

Fact 1: \emph{There is a one-to-one correspondence between flat D-branes
in $X$, modulo Baum-Douglas equivalence, and stable homotopy classes
of Fredholm modules over the algebra $C(X)$.}

In the presence of a non-zero $B$-field on $X$, which is a $U(1)$-gerbe
with connection represented by the characteristic class in $H^{3}(X,\mathbb{Z})$
\cite{Szabo2008b,AsselmeyerKrol2009}, one can define twisted D-brane
on $X$ as \cite{Szabo2008b}:

\begin{definition}

A twisted D-brane in a B-field $(X,H)$ is a triple $(W,E,\phi)$,
where $\phi:W\hookrightarrow X$ is a closed, embedded oriented submanifold
with $\phi^{\star}H={\rm W}_{3}(W)$, and $E$ is the Chan-Paton bundle
on $W$, i.e. $E\in K^{0}(W)$, and ${\rm W}_{3}(W)$ is the 3-rd
integer Stiefel-Whitney class of the normal bundle of $W$, ${\rm W}_{3}(W)\in H^{3}(W,\mathbb{Z})$. 

\end{definition}

The condition in the definition is in fact required by the cancellation
of the Freed-Witten anomaly, where $H\in H^{3}(X,\mathbb{Z})$ is
the NS-NS $H$-flux. Since ${\rm W}_{3}(W)$ is the obstruction to
the ${\rm spin}^{c}$ structure on $W$, in the case of ${\rm W}_{3}(W)=0$
one has flat D-branes in $X$. Thus equivalence classes of twisted
D-branes on $X$ are represented by twisted topological K-homology
$K_{\star}(X,H)$ which is dual to the twisted K-theory $K^{\star}(X,H)$.
As was argued in \cite{AsselmeyerKrol2010}, in case of $S^{3}$,
one has some exotic $\mathbb{R}^{4}$'s which can be twisted by $H$
leading to the K-theory $K^{\star}(S^{3},H)$. We can represent the
$U(1)$ gerbes with connection on $S^{3}$, by the bundles ${\cal E}_{H}$
of algebras over $S^{3}$, such that the sections of the bundle ${\cal E}_{H}$
define the noncommutative, twisted algebra \emph{$C_{0}(X,{\cal E}_{H})$
}and the Dixmier-Douady class of ${\cal E}_{H}$, $\delta_{H}({\cal E}_{H})$,
is $H\in H^{3}(S^{3},\mathbb{Z})$ \cite{AsselmeyerKrol2009a,AtiyahSegal2004,Szabo2002a}.
The important relation is the following (\cite{Szabo2008b}, Proposition
1.15): 

Fact 2: \emph{There is a one-to-one correspondence between twisted
D-branes in $(X,H)$ and stable homotopy classes of Fredholm modules
over the algebra $C_{0}(X,{\cal E}_{H})$.}

Since the algebra \emph{$C_{0}(X,{\cal E}_{H})$ }certainly determines
its stable homotopy classes of the Fredholm modules on it, then in
the case $X=S^{3}$ one has the following observation:

A. \emph{Let the exotic smooth $\mathbb{R}^{4}$'s are determined
by the integral third classes $H\in H^{3}(S^{3},\mathbb{Z})$. Then,
these exotic smooth $\mathbb{R}^{4}$'s correspond one-to-one to the
set of twisted D-branes in $(S^{3},H)$.}

In principle, given the complete collection of twisted D-branes in
$(S^{3},H)$, which take values in $K_{\star}(S^{3},H)$, one can
determine the corresponding exotic $\mathbb{R}^{4}$. This is simply
the exotic $\mathbb{R}_{H}^{4}$ corresponding to the class $[H]\in H^{3}(S^{3})$
and $H$ makes the twist in the K-homology as dual to the twisted
K-theory $K^{\star}(S^{3},H)$ \cite{AsselmeyerKrol2009a,AsselmeyerKrol2010,Szabo2002a}.
In this paper we collect further evidences that this is also the case
more generally, and the relation D-branes - 4-exotics is closer. 

Remembering that $S^{3}\subset\mathbb{R}^{4}$ as part of the Akbulut
cork of the exotic structure, our previous observation can be restated
as:

B. \emph{The change of the exotic smoothness of $\mathbb{R}^{4}$,
$\mathbb{R}_{H_{1}}^{4}\to\mathbb{R}_{H_{2}}^{4}$, $H_{1}$, $H_{2}\in H^{3}(S^{3},\mathbb{Z})$,
$H_{1}\neq H_{2}$, corresponds to the change of the curved backgrounds
$(S^{3},H_{1})\to(S^{3},H_{2})$ hence the sets of stable D-branes.}

This motivates the formulation:

C. \emph{Some small exotic smoothness on $\mathbb{R}^{4}$, $\mathbb{R}_{H_{1}}^{4}$,
can be destabilize (or stabilize) D-branes in $(S^{3},H_{2})$, where
$S^{3}\subset\mathbb{R}^{4}$ lies at the boundary of the Akbulut
cork of $\mathbb{R}_{H_{1}}^{4}$. We say that D-branes in $(S^{3},H_{2})$
are }4-exotic-sensitive\emph{.}

Turning to the generalization of spaces to noncommutative $C^{\star}$
algebras, there were developed recently impressive counterparts of
many topological, geometrical and analytical results, like Poincar\'e
duality, characteristic classes and the Riemann-Roch theorem. Also
the generalized formula for charges of quantum D-branes in a noncommutative
separable $C^{\star}$ algebras was worked out \cite{Szabo2008a,Szabo2008b}.
Thus the suitable framework for considering the quantum regime of
D-branes emerged. In next subsection we will try to find a relation
to 4-exotics also in this quantum regime of D-branes. 

Following \cite{AsakawaSugimotoTerasima2002,Szabo2008a,Szabo2008b,Szabo2008c}
one can take as an initial substitute for the category of quantum
D-branes, the category of separable $C^{\star}$ algebras and morphisms
being elements of KK theory groups. This means that for a pair $({\cal A},{\cal B})$
of separable $C^{\star}$ algebras the morphisms $h:{\cal A}\to{\cal B}$
is lifted to the element of the group $KK({\cal A},{\cal B})$. Thus
we can consider a generalized D-branes in a separable $C^{\star}$
algebra ${\cal A}$ as corresponding to the lift $h!:{\cal A}\to{\cal B}$
where ${\cal B}$ represents a quantum D-brane.

More precisely following \cite{Szabo2008a}, let us consider a subcategory
${\cal C}$ of the category of $C^{\star}$ separable algebras and
their morphisms, which consists of strongly K-oriented morphisms.
This means that there exists a contravariant functor $!:{\cal C}\to KK$
such that ${\cal C}\ni f:{\cal A}\to{\cal B}$ is mapped to $f!\in KK_{d}({\cal B},{\cal A})$,
here $KK$ is the category of separable $C^{\star}$ algebras with
KK classes as morphisms. Strongly K-oriented morphisms and the functor
$!$ are subjects to the following conditions:

\begin{enumerate}
\item Identity morphism $id_{{\cal A}}:{\cal A}\to{\cal A}$ is strongly
K-oriented (SKKO) for every separable $C^{\star}$ algebra ${\cal A}$
and $(id_{{\cal A}})!=1_{{\cal A}}$. Also, the 0-morphism $0_{{\cal A}}:{\cal A}\to{\cal A}$
is SKKO and $(0_{{\cal A}})!=0\in KK(0,{\cal A})$.
\item If $f:{\cal A}\to{\cal B}$ is SKKO then $f^{\circ}:{\cal A}^{\circ}\to{\cal B}^{\circ}$
is either, and $(f!)^{\circ}=(f^{\circ})!$. ${\cal A}^{\circ}$ is
the opposite $C^{\star}$ algebra to ${\cal A}$, i.e. one which has
the same underlying vector space but reversed product. 
\item Any morphism $f:{\cal A}\to{\cal B}$ is SKKO, provided ${\cal A}$
and ${\cal B}$ are strong Poincar\'e dual (PD) algebras. Then $f!$
is determined as: \begin{equation}
f!=(-1)^{d_{{\cal A}}}\Delta_{{\cal A}}^{\vee}\otimes_{{\cal A}^{0}}\left[f^{0}\right]\otimes_{{\cal B}^{0}}\Delta_{{\cal B}}\label{eq:K-orientation}\end{equation}
here $[f]$ is the class of $f:{\cal A}\to{\cal B}$ in $KK({\cal A},{\cal B})$.
$\Delta_{{\cal A}}$ is the fundamental class in $KK_{d_{{\cal A}}}({\cal A}\otimes{\cal A}^{\circ},\mathbb{C})=K^{d_{{\cal A}}}({\cal A}\otimes{\cal A}^{\circ})$,
$\Delta_{{\cal A}}^{\vee}$ its dual class in $KK_{-d_{{\cal A}}}(\mathbb{C},{\cal A}\otimes{\cal A}^{\circ})=K_{-d_{{\cal A}}}(A\otimes{\cal A}^{\circ})$
which exist by strong PD \cite{Szabo2008a}. 
\end{enumerate}
K-orientability was introduced, in its original form, by A. Connes
in order to define the analogue of ${\rm spin}^{c}$ structure for
noncommutative $C^{\star}$ algebras (see also \cite{Connes1984}
and next subsections). Presented here formulation of K-orientability
and strong PD $C^{\star}$ algebras are crucial ingredients of noncommutative
versions of Riemann-Roch theorem, Poincar\'e-like dualities, Gysin
K-theory map and allows to formulate a very general formula for noncommutative
D-brane charges \cite{Szabo2008b,Szabo2008a,Szabo2008c}. Let us notice
that if both ${\cal A}$ and ${\cal B}$ are PD algebras then any
morphism $f:{\cal A}\to{\cal B}$ is K-oriented and the K-orientation
for $f$ is given in (\ref{eq:K-orientation}). 

In the particular case of the proper smooth embedding $f:M\to X$
of codimension $d$, where $M$, $X$ are smooth compact manifolds,
let the normal bundle $\tau$ over $W$, of $TW$ with respect to
$f^{\star}(TX)$, be ${\rm spin}^{c}$. When also $X$ is ${\rm spin}^{c}$
then the ${\rm spin}^{c}$ condition on $\tau$ when $H$-flux is
absent in type II string theory formulated on $X$, is the Freed-Witten
anomaly cancellation condition \cite{Szabo2008a}. In this case any
D-brane in $X$, given by the triple $(W,E,f)$, determines the KK-theory
element $f!\in KK(C(W),C(X))$. The construction of K-orientation
$f:M\to X$, between smooth compact manifolds, can be extended to
smooth proper maps which are not necessary embeddings. Thus the general
condition for K-orientability gives the correct analogue for stable
D-branes in $C^{\star}$ algebras. 

\begin{definition}\label{enu:Def: q-Branes}

\emph{A generalized stable quantum D-brane} on a separable $C^{\star}$
algebra ${\cal A}$, represented by a separable $C^{\star}$ algebra
${\cal B}$, is given by the strongly K-oriented homomorphism of $C^{\star}$
algebras, $h_{{\cal B}}:{\cal A}\to{\cal B}$. The K-orientation means
that there is the lift $(h_{{\cal B}})!\in KK({\cal B},{\cal A})$
where $!$ fulfills the functoriality condition as in (\ref{eq:K-orientation}).

\end{definition}

This kind of an approach to quantum D-branes is in fact a conjectural
framework which exceeds both the dynamical Seiberg-Witten limit of
superstring theory (where noncommutative brane world-volumes emerges)
and geometrical understanding of branes, and places itself rather
in a deep quantum regime of the theory \cite{Szabo2008c}.

\subsection{Exotic $\mathbb{R}^{4}$ and stable D-branes configurations on foliated
manifolds\label{sub:D-branes-on-foliated}}

Now we want to approach the problem of description of stable states
of D-branes in a more general geometry than spaces, namely the geometry
of foliated manifolds. The case of our interest is a codimension-1
foliation of $S^{3}$. This is a noncommutative geometry. In general,
to every foliation $(V,F)$ one can associate its noncommutative $C^{\star}$
algebra $C^{\star}(V,F)$, on the other hand a foliation determines
its holonomy groupoid $G$ and the topological classifying space $BG$.
Both cases, topological K-homology of $G$ and $C^{\star}$algebraic
K-theory, are in fact dual. Analogously to our previous discussion
of branes as K-cycles on $X$, let us start with K-homology of $G$
and define D-branes as K-cycles in $G$:

A $K$ - cycle on a foliated geometry $X=(V,F)$ is a triple $(M,E,\phi)$
where $M$ is a compact manifold without boundary, $E$ is a complex
vector bundle on $M$ and $\phi:M\to BG$ is a smooth K-oriented map.
Due to the K-orientability in the presence of canonical $G$-bundle
$\tau$ on $BG$, the condition of ${\rm Spin}^{c}$ structure on
$M$ is lifted to the ${\rm Spin}^{c}$ structure on $TM\oplus\phi^{\star}\tau$
\cite{Connes1984}. 

The topological $K$-homology $K_{\star,\tau}(X)=K_{\star,\tau}(BG)$
of the foliation $(V,F)$ is the set of equivalence classes of the
above triples, where the equivalence respects the following conditions:

\begin{itemize}

\item[(i)] $(M_{1},E_{1},\phi_{1})\sim(M_{2},E_{2},\phi_{2})$ when
there exists a triple (bordism of the triples) $(M,E,\phi)$ such
that $(\partial M,E_{|\partial M},\phi_{|\partial M})$ is isomorphic
to the disjoint union $(M_{1},E_{1},\phi_{1})\cup(-M_{2},E_{2},\phi_{2})$
where $-M_{2}$ is the reversed ${\rm {Spin}^{c}}$ structure of $TM_{2}\oplus\phi_{2}^{\star}\tau$
and $M$ is a compact manifold with boundary. 

\item[(ii)] $(M,E_{1}\oplus E_{2},\phi)\sim(M,E_{1},\phi)\cup(M,E_{2},\phi)$,

\item[(iii)] Vector bundle modification $(M,E,\phi)\sim(\widehat{M},\widehat{H}\otimes\rho^{\star}(E),\phi\circ\rho)$
similarly as in the case of manifolds. 

\end{itemize}

As in the case of spaces (manifolds) and the corresponding K-homology
groups representing stable D-branes of type II superstring theory
(see Sec. \ref{sub:D-branes-on-spaces:}), also here, in the case
of the geometry of foliated manifolds we generalize stable D-branes
as being represented by the above triples. 

\begin{theorem}

The class of generalized stable D-branes on the $C^{\star}$ algebra
$C^{\star}(S^{3},F_{1})$ (of the codimension 1 foliation of $S^{3}$)
which correspond to the K-homology classes $K_{\star,\tau}(S^{3}/F)$,
determines an invariant of exotic smooth $\mathbb{R}^{4}$. Such an
exotic $\mathbb{R}^{4}$ contains this foliated $S^{3}$ as a generalized
(noncommutative) smooth subset \cite{AsselmeyerKrol2009a}. 

\end{theorem}

The result follows from the fact that $K_{\star,\tau}(S^{3}/F)$ is
isomorphic to $K_{\star,\tau}(BG)$ \cite{Connes1984} and this determines
a class of stable D-branes in $(S^{3},F)$. The foliations $(S^{3},F)$
correspond to different smoothings on $\mathbb{R}^{4}$ \cite{AsselmeyerKrol2009}.
$\square$ 

Let us note that this approach allows for considering a kind of string
theory and branes also beyond the integral levels of $SU(2)$ WZW
model given by $[H]\in H^{3}(S^{3},\mathbb{Z})$. The relation with
exotic smooth $\mathbb{R}^{4}$'s extends over this as well.

\subsection{Net of exotic $\mathbb{R}^{4}$'s and quantum D-branes in $C^{\star}(S^{3},F)$\label{sub:Net-of-exotic}}

The extension of string theory and D-branes over general noncommutative
separable $C^{\star}$ algebras where also D-branes are represented
by noncommutative separable $C^{\star}$ algebras, can be considered
as an approach to quantum D-branes. A category of D-branes in a quantum
regime, is the category of separable $C^{\star}$ algebras and morphisms
which are elements of KK theory groups. For a pair $({\cal A},{\cal B})$
of separable $C^{\star}$ algebras the morphisms $h:{\cal A}\to{\cal B}$
belong to $KK({\cal A},{\cal B})$. Abstract quantum D-branes in a
separable $C^{\star}$ algebra ${\cal A}$ correspond to $\phi:{\cal A}\to{\cal B}$
where ${\cal B}$ is the algebra representing a quantum D-brane and
$\phi$ is a strongly K-oriented map. For such branes a general formula
for RR charges in noncommutative setting was worked out \cite{Szabo2008a,Szabo2008b}.

D-branes considered in the previous subsection, correspond to the
lifted KK-theory classes, i.e. $f!\in KK(M,V/F)$ where D-brane corresponds
to the triple $(M,E,f)$ and $f:M\hookrightarrow G=V/F$ is K-oriented
map (see \cite{Connes1984}). More generally (still following \cite{Connes1984}),
given a K-oriented map $f:X\to Y$ , one can define (under certain
conditions) a push forward map $f!$ in K-theory. The very important
property of the analytical group $K(V/F)$ of the foliation $(V,F)$
is its ,,wrong way'' (Gysin) functoriality which to each K-oriented
map $f:V_{1}/F_{1}\to V_{2}/F_{2}$ of leaf spaces associates an element
$f!$ of the Kasparov group $KK(C^{\star}(V_{1};F_{1});C^{\star}(V_{2};F_{2}))$. 

Now given a small exotic $\mathbb{R}^{4}$, say $e_{1}$, embedded
in some small exotic $\mathbb{R}^{4}$, $e$, both are represented
by the $C^{\star}$ algebras of the codimension-1 foliations of $S^{3}$,
$C^{\star}(V_{1};F_{1})$ and $C^{\star}(V;F)$ respectively. The
embedding $i:e_{1}\hookrightarrow e$ determines the corresponding
K-oriented map of the leaf spaces $f_{i}:S^{3}/F_{1}\to S^{3}/F$
and the KK-theory lift $f_{i}!:KK(C^{\star}(V_{1};F_{1});C^{\star}(V;F))$.
According to Def. \ref{enu:Def: q-Branes} from Sec. \ref{sub:Branes-on-separable},
we see that

\begin{theorem}

\label{theo:quantum-exotic-R4}

Let $e$ be an exotic $\mathbb{R}^{4}$ corresponding to the codimension-1
foliation of $S^{3}$ which gives rise to the $C^{\star}$algebra
${\cal A}_{e}$. The exotic smooth $\mathbb{R}^{4}$ embedded in $e$
determines a generalized quantum D-brane in ${\cal A}_{e}$. 

\end{theorem}

Given exotic $\mathbb{R}^{4}$'s, $\{e_{a},\, a\in I\}$, all embedded
in $e$, one has the family of $C^{\star}$ algebras, $\{{\cal A}_{a},\, a\in I\}$,
of the codimension-1 foliations of $S_{a}^{3},\: a\in I$. Now the
embeddings $e_{a}\to e$ determine the corresponding K-oriented maps
of the leaf spaces as before, and the $\star$-homomorphisms of algebras
$\phi_{a}:{\cal A}_{e}\to{\cal A}_{a}$. The corresponding classes
in KK theory $KK({\cal A}_{e},{\cal A}_{a})$, represent quantum D-branes
in ${\cal A}_{e}$. $\square$

However, the correspondence in the theorem is many-to-one and an exotic
smooth $\mathbb{R}^{4}$ embedded in $e$ can be represented (non-uniquely)
by stable D-brane in ${\cal A}_{e}$, and not all abstract D-branes
in the algebra ${\cal A}_{e}$ are represented by some exotic $e'\subset e$.
Still one can consider D-branes represented by exotic $e_{a}$ in
$e$ as carrying 4-dimensional, hence potentially physical, information.
This is a kind of special ,,superselection'' rule in superstring theory
and will be discussed separately.

\subsection{RR charges of D6-Branes in the presence of $B$-field}

Now let us comment on some indication how 4-dimensional structure
can refer directly to dynamics of higher dimensional branes in flat
spacetime. This higher dimensional brane is the important D6-brane
which is usually involved in building various ,,realistic'' 4-dimensional
models derived from the brane configurations. We will analyze this
case separately along with compactifications in string theory.

Let us consider the D6-brane of IIA string theory in flat 10 dimensional
spacetime and assume that B-field vanishes. The world-volumes of flat
Dp-branes are classified by $K_{1}(\mathbb{R}^{p+1})$ where this
K-homology group is understood as $K^{1}(C_{0}(\mathbb{R}^{p+1}))$
i.e. the K-group of the reduced $C^{\star}$ algebra of functions
$C_{0}(\mathbb{R}^{p+1})=C(S^{p+1})$. Hence $K_{1}(\mathbb{R}^{p+1})=K_{1}(S^{p+1})$.
Their charges, constraining the dynamics of the brane, are dually
described by $K^{1}(\mathbb{R}^{9-p})=K^{1}(S^{9-p})$. In the case
of D6-branes we have $K^{1}(S^{3})$ as classifying the RR charges
of flat D6-branes in flat 10-dimensional spacetime \cite{Witten1998}. 

In the presence of a non-vanishing B-field for a stable D6-brane,
the B-field need to be non-trivial on space $\mathbb{R}^{3}$ transverse
to the world-volume , hence $S^{3}$. The classification of D6-brane
charges in IIA type superstring theory in flat space is then given
by the twisted K-theory $K_{H}(S^{3})$, which is $K^{1}(S^{3},H)=\mathbb{Z}_{k}$,
where $0\neq[dB]=[H]=k\in H^{3}(S^{3},\mathbb{Z})$. Hence the dynamics
of D6-branes in type IIA superstring theory on flat spacetime is influenced
by non-zero B-field. 

Now we follow the philosophy present already implicitly in our previous
work that the source for the non-trivial B-field on $S^{3}$, hence
$H\neq0$, is due to the exoticness of the ambient $\mathbb{R}^{4}$.
The motivation is certainly the fact that the given exotic $\mathbb{R}_{H}^{4}$
corresponds to the non-trivial class $[H]\in H^{3}(S^{3},\mathbb{Z})$
and conversely, where $S^{3}$ is taken from the boundary of the Akbulut
cork \cite{AsselmeyerKrol2009,AsselmeyerKrol2009a}. Moreover, exotic
smoothness of $\mathbb{R}_{H}^{4}$ twists the K-theory groups $K^{\star}(S^{3})$
\cite{AsselmeyerKrol2010} provided $S^{3}$ lies at the boundary
of the Akbulut cork.

Hence the possible dynamics (the charges) of D6-branes in spacetime
$\mathbb{R}_{H}^{4}\times\mathbb{R}^{5,1}$ is equivalently referred
to the dynamics of D6-brane in the presence of non-zero B-field on
transversal $\mathbb{R}^{3}$. 

\begin{theorem}

RR charges of D6-branes in string theory IIA in the presence of non-trivial
B-field ($[dB]\neq0$), (these charges are classified by $K_{H}(S^{3})$,
$H\neq0$ and $[H]\in H^{3}(S^{3},\mathbb{Z})$), are related with
exotic smoothness of small $\mathbb{R}_{H}^{4}$. This exotic $\mathbb{R}_{H}^{4}$
corresponds to $[H]$ which twists $K(S^{3})$ \cite{AsselmeyerKrol2010},
where $S^{3}\subset\mathbb{R}^{4}$ lies at the boundary of the Akbulut
cork and $S^{3}$ is transverse to the branes. Thus, changing the
smoothness of $\mathbb{R}^{4}$ gives rise to the change of the allowed
charges for D6 branes, hence the dynamics changes. 

\end{theorem}

We see that geometric realization of (classical) D-branes in certain
backgrounds of string theory is correlated with small exotic $\mathbb{R}^{4}$'s
which can be all embedded in the standard smooth $\mathbb{R}^{4}$.
We saw in previous subsection \ref{sub:Net-of-exotic} that quantum
D-branes correspond to the net of exotic smooth $\mathbb{R}^{4}$'s
embedded in certain exotic smooth $\mathbb{R}^{4}$. Also an intriguing
interpretation for this correspondence can be given: \emph{in some
limit of IIA superstring theory, small exotic smooth $\mathbb{R}^{4}$'s
can be considered as carrying the RR charges of D6 branes}. 

We will come back to these interesting points in the next section.

\section{From wild embeddings to quantum D-branes}

In this section we try to give a geometric approach to quantum D-branes
using wild embeddings of trivial complexes into $S^{n}$ or $\mathbb{R}^{n}$.
Furthermore we are able to obtain a low-dimensional interpretation
of D-brane charges. This point of view is supported by the Theorem
\ref{theo:quantum-exotic-R4} above. Here we will describe a dimension-independent
way: every wild embedding $j$ of a $p-$dimensional complex $K$
into the $n-$dimensional sphere $S^{n}$ is determined by the fundamental
group $\pi_{1}(S^{n}\setminus j(K))$ of the complement. This group
is perfect and uniquely representable by a 2-dimensional complex,
a singular disk or grope (see \cite{Can:79}). As we showed in \cite{AsselmeyerKrol2009},
the exotic $\mathbb{R}^{4}$ is given by the grope. Thus, every quantum
D-brane must be determined (as a kind of germ) by some exotic $\mathbb{R}^{4}$.

\subsection{Wild and tame embeddings\label{sub:Wild-and-tame-embed}}

We call a map $f:N\to M$ between two topological manifolds an embedding
if $N$ and $f(N)\subset M$ are homeomorphic to each other. From
the differential-topological point of view, an embedding is a map
$f:N\to M$ with injective differential on each point (an immersion)
and $N$ is diffeomorphic to $f(N)\subset M$. An embedding $i:N\hookrightarrow M$
is \emph{tame} if $i(N)$ is represented by a finite polyhedron homeomorphic
to $N$. Otherwise we call the embedding \emph{wild}. There are famous
wild embeddings like Alexanders horned sphere or Antoine's necklace.
In physics one uses mostly tame embeddings but as Cannon mentioned
in his overview \cite{Can:78}, one needs wild embeddings to understand
the tame one. As shown by us \cite{AsselmeyerKrol2009}, wild embeddings
are needed to understand exotic smoothness. As explained in \cite{Can:78}
by Cannon, tameness is strongly connected to another topic: decomposition
theory (see the book \cite{Daverman1986}). 

Two embeddings $f,g:N\to M$ are said to be isotopic, if there exists
a homeomorphism $F:M\times[0,1]\to M\times[0,1]$ such that 

\begin{enumerate}
\item $F(y,0)=(y,0)$ for each $y\in M$ (i.e. $F(.,0)=id_{M}$)
\item $F(f(x),1)=g(x)$ for each $x\in N$, and
\item $F(M\times\left\{ t\right\} )=M\times\left\{ t\right\} $ for each
$t\in[0,1]$.
\end{enumerate}
If only the first conditions can be fulfilled then one call it concordance.
Embeddings are usually classified by isotopy. An important example
is the embedding $S^{1}\to\mathbb{R}^{3}$, known as knot, where different
knots are different isotopy classes.

\subsection{Embeddings of $(4k-1)$- into $6k$-manifolds}

Now we start with a short discussion of embeddings $S^{3}\to S^{6}$
as the example $k=1$ of a general map $S^{4k-1}\to S^{6k}$. As Haefliger
\cite{Haefliger1962} showed the isotopy classes of embeddings are
determined by the integer classes (Hopf invariant) in $H^{3}(S^{3},\mathbb{Z})$.
Thus the $4k-1$ space is knotted in the $6k$ space. This phenomenon
depends strongly on smoothness, i.e. it disappears for continuous
or PL embeddings. Usually every $n-$sphere or every homology $n-$sphere
unknots (in PL or TOP) in $\mathbb{R}^{m}$ for $m\geq n+3$, i.e.
for codimension $m-n=3$ or higher. Of course, one has the usual knotting
phenomena in codimension $2$ and the codimension $1$ was shown to
be unique for embeddings $S^{n}\to S^{n+1}$ (for $n\geq6$) but is
hard to solve in other cases. 

Let $\Sigma\to S^{6}$ be an embedding of a homology 3-sphere $\Sigma$
(containing the case $S^{3}$). Then the normal bundle of $F$ is
trivial (definition of an embedding) and homotopy classes of trivializations
of the normal bundle (normal framing) are classified by the homotopy
class $[\Sigma,SO(3)]$ with respect to some fixed framing. There
is an isomorphism $[\Sigma,SO(3)]=[\Sigma,S^{2}]$ (so-called Pontrjagin-Thom
construction) and $[\Sigma,S^{2}]$ can be identified with $H^{3}(\Sigma,\mathbb{Z})=\mathbb{Z}$.
That is one possible way to get the classification of isotopy classes
of embeddings $\Sigma\to S^{6}$ by elements of $H^{3}(\Sigma,\mathbb{Z})=\mathbb{Z}$.
A class $[H]$ in $H^{3}(\Sigma,\mathbb{Z})$ determines via Stokes
theorem\[
\intop_{\Sigma=\partial A}H=\intop_{A}dH\]
the 4-form $dH$ in the 4-manifold%
\footnote{To every 3-manifold $\Sigma$, there is a 4-manifold $A$ with $\partial A=\Sigma$.%
} $A$ with $\partial A=\Sigma$. As we know the (small) exotic $\mathbb{R}^{4}$
is determined by a contractable submanifold $A$, the Akbulut cork,
with boundary $\partial A$ a homology 3-sphere. The contractability
of $A$ implies $H^{4}(A,\mathbb{Z})=0$, i.e. every 4-form on $A$
is given by $dH$ for some 3-form $H$. The isomorphism $H^{4}(A,\partial A)=H^{3}(\partial A)$
and Stokes theorem imply\[
\intop_{A}dH=\intop_{\partial A}H=Q\not=0\]
the non-vanishing of the 4-form $dH=Q\cdot dvol(A)$ with the volume
form $dvol(A)$ of $A$ normed to one. Combined with our result that
$H^{3}(S^{3},\mathbb{Z})$ determines some exotic $\mathbb{R}^{4}$
we have:

\begin{theorem}(The topological origins of the allowed D6-brane charges)

\label{theo:D6-brane-charge}

Let $\mathbb{R}_{H}^{4}$ be some exotic $\mathbb{R}^{4}$ determined
by some 3-form $H$, i.e. by a codimension-1 foliation on the boundary
$\partial A$ of the Akbulut cork $A$. The codimension-1 foliation
on $\partial A$ is determined by $H^{3}(\partial A,\mathbb{R})$.
Each integer class in $H^{3}(\partial A,\mathbb{Z})$ determines the
isotopy class of an embedding $\partial A\to S^{6}$. Hence, the group
of allowed charges of D6-branes in the presence of B-field in $M^{10}$,
i.e. $K_{H}^{3}(S^{3})$ with $dB=H$, is determined equivalently
by the isotopy classes of embeddings $\partial A\to S^{6}$. The classes
of $H$-field are topologically determined by the isotopy classes
of the embeddings, which affects the allowed charges of D6-branes. 

\end{theorem}

But more is true. Given two embeddings $F_{i}:\Sigma_{i}\to S^{6}$
between two homology 3-spheres $\Sigma_{i}$ for $i=0,1$. A homology
cobordism is a cobordism between $\Sigma_{0}$ and $\Sigma_{1}$.
This cobordism can be embedded in $S^{6}\times[0,1]$ determining
the homology bordism class of the embedding. Then two embeddings of
an oriented homology 3-sphere in $S^{6}$ are isotopic if and only
if they are homology bordant.

\subsection{Real cohomology classes and wild embeddings}

Wild embeddings are important to understand usual embeddings. Consider
a closed curve in the plane. By common sense, this curve divides the
plane into an interior and an exterior area. The Jordan curve theorem
agrees with that view completely. But what about one dimension higher,
i.e. consider the embedding $S^{2}\to\mathbb{R}^{3}$? Alexander was
the first who constructed a counterexample, Alexanders horned sphere
\cite{Alex:24}, as wild embedding $D^{2}\to\mathbb{R}^{3}$. The
main property of this wild object $D_{W}^{2}$ is the non-simple connected
complement $\mathbb{R}^{3}\setminus D_{W}^{2}$. In the following
we will concentrate on wild embeddings of spheres $S^{n}$ into spheres
$S^{m}$ equivalent to embeddings of $\mathbb{R}^{n}$ into $\mathbb{R}^{m}$relative
to the infinity $\infty$ point or to relative embeddings of $D^{n}$
into $D^{m}$ (relative to its boundary). From the physical point
of view, D-branes or M-branes are topological objects of a trivial
type like $\mathbb{R}^{n},S^{n}$ or $D^{n}$. 

Lets start with the case of a finite $k-$dimensional polyhedron $K^{k}$
(i.e. a piecewise-linear version of a $k-$disk $D^{k}$). Consider
the wild embedding $i:K\to S^{n}$ with $0\leq k\leq n-3$ and $n\geq7$.
Then, as proofed in \cite{FerryPedersenVogel1989}, the complement
$S^{n}\setminus i(K)$ is non-simple connected with a countable generated
(but not finitely presented) fundamental group $\pi_{1}(S^{n}\setminus i(K))=\pi$.
Furthermore, the group $\pi$ is perfect (i.e. generated by the commutator
subgroup $[\pi,\pi]=\pi$ implying $H_{1}(\pi)=0$) and $H_{2}(\pi)=0$
($\pi$ is called a superperfect group). With other words, $\pi$
is a group where every element $x\in\pi$ can be generated by a commutator
$x=[a,b]=aba^{-1}b^{-1}$ (including the trivial case $x=a,\: b=e$).
By using geometric group theory, we can represent $\pi$ by a grope
(or generalized disk, see Cannon \cite{Can:79}), i.e. a hierarchical
object with the same fundamental group as $\pi$ (see below). In \cite{AsselmeyerKrol2009},
the grope was used to construct a non-trivial involution of the 3-sphere
connected with a codimension-1 foliation of the 3-sphere classified
by the real cohomology classes $H^{3}(S^{3},\mathbb{R})$. By using
the suspension \[
\Sigma X=X\times[0,1]/(X\times\left\{ 0\right\} \cup X\times\left\{ 1\right\} \cup\left\{ x_{0}\right\} \times[0,1])\]
of a topological space $(X,x_{0})$ with base point $x_{0}$, we have
an isomorphism of cohomology groups $H^{n}(S^{n})=H^{n+1}(\Sigma S^{n})$.
Thus the class in $H^{3}(S^{3},\mathbb{R})$ induces classes in $H^{n}(S^{n},\mathbb{R})$
for $n>3$ represented by a wild embedding $i:K\to S^{n}$ for some
$k-$dimensional polyhedron. Then every small exotic $\mathbb{R}^{4}$
determines also higher brane charges:

\begin{theorem}

Let $\mathbb{R}_{H}^{4}$ be some exotic $\mathbb{R}^{4}$ determined
by element in $H^{3}(S^{3},\mathbb{R})$, i.e. by a codimension-1
foliation on the boundary $\partial A$ of the Akbulut cork $A$.
Each wild embedding $i:K^{3}\to S^{p}$ for $p>6$ of a 3-dimensional
polyhedron (as part of $S^{3}$) determines a class in $H^{p}(S^{p},\mathbb{R})$
which can be interpreted as the charge of a $Dp$ brane in the sense
of Theorem \ref{theo:D6-brane-charge}.

\end{theorem}

\subsection{$C^{*}-$algebras associated to wild embeddings}

\begin{figure}
\begin{center}

\includegraphics[height=8cm]{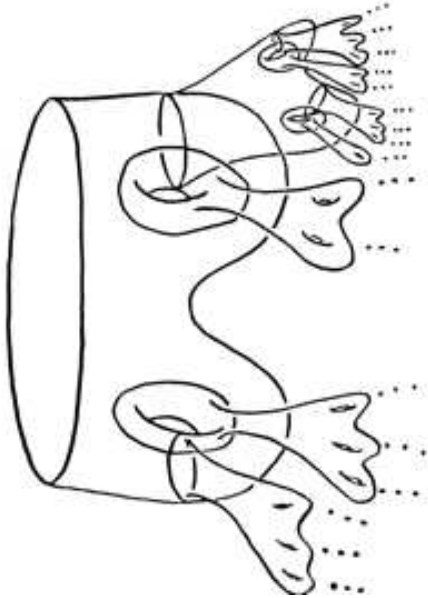}

\end{center}\caption{An example of a grope\label{fig:grope}}

\end{figure}

As described above, a wild embedding $j:K\to S^{n}$ of a polyhedron
$K$ is characterized by its complement $M(K,j)=S^{n}\setminus j(K)$
which is non-simple connected (i.e. the fundamental group $\pi_{1}(M(K,j))$
is non-trivial). The fundamental group $\pi_{1}(M(K,j))=\pi$ of the
complement $M(K,j)$ is a superperfect group, i.e. $\pi$ is identical
to its commutator subgroup $\pi=[\pi,\pi]$ (then $H_{1}(\pi)=0)$
and $H_{2}(\pi)=0$. This group is not finite in case of a wild embedding.
Here we use gropes to represent $\pi$ geometrically. The idea behind
that approach is very simple: the fundamental group of the 2-dimensional
torus $T^{2}$ is the abelian group $\pi_{1}(T^{2})=\left\langle a,b\:|\:[a,b]=aba^{-1}b^{-1}=e\right\rangle =\mathbb{Z}\oplus\mathbb{Z}$
generated by the two standard slopes $a,b$. The capped torus $T^{2}\setminus D^{2}$
has an additional element $c$ in the fundamental group generated
by the boundary $\partial(T^{2}\setminus D^{2})=S^{1}$. This element
is represented by the commutator $c=[a,b]$. In our superperfect group
we have the same problem: every element $c$ is generated by the commutator
$[a,b]$ of two other elements $a,b$ which are also represented by
commutators etc. Thus one obtains a hierarchical object, a generalized
2-disk or a grope (see Fig. \ref{fig:grope}).%

Now we describe two ways to associate a $C^{*}-$algebra to this grope.
This first approach uses a combination of our previous papers \cite{AsselmeyerKrol2009,AsselmeyerKrol2010}.
Then every grope determines a codimension-1 foliation of the 3-sphere
and vice verse. The leaf-space of this foliation is a factor $I\! I\! I_{1}$
von Neumann algebra and we have a $C^{*}-$algebra for the holonomy
groupoid. For later usage, we need a more direct way to construct
a $C^{*}-$algebra from a wild embedding or grope. The main ingredient
is the superperfect group $\pi$, countable generated but not finitely
presented group. To get an impression of this group, we consider a
representation $\pi\to G$ in some infinite group. As the obvious
example for $G$ we choose the infinite union $GL(\mathbb{C})=\bigcup_{\infty}GL(n,\mathbb{C})$
of complex, linear groups (induced from the embedding $GL(n,\mathbb{C})\to GL(n+1,\mathbb{C})$
by an inductive limes process). Then we have a homomorphism\[
U:\pi\to GL(\mathbb{C})\]
mapping a commutator $[a,b]\in\pi$ to $U([a,b])\in[GL(\mathbb{C}),GL(\mathbb{C})]$
into the commutator subgroup of $GL(\mathbb{C})$. But every element
in $\pi$ is generated by a commutator, i.e. we have\[
U:\pi\to[GL(\mathbb{C}),GL(\mathbb{C})]\]
and we are faced with the problem to determine this commutator subgroup.
Actually, one has Whitehead's lemma (see \cite{Ros:94}) which determines
this subgroup to be the group of elementary matrices $E(\mathbb{C})$.
One defines the elementary matrix $e_{ij}(a)$ in $E(n,\mathbb{C})$
to be the $(n\times n)$ matrix with 1\textasciiacute{}s on the diagonal,
with the complex number $a\in\mathbb{C}$ in the $(i,j)-$slot, and
0\textasciiacute{}s elsewhere. Analogously, $E(\mathbb{C})$ is the
infinite union $E(\mathbb{C})=\bigcup_{\infty}E(n,\mathbb{C})$. Thus,
every homomorphism descends to a homomorphism\[
U:\pi\to E(\mathbb{C})=[GL(\mathbb{C}),GL(\mathbb{C})]\quad.\]
By using the relation\[
[e_{ij}(a),e_{jk}(b)]=e_{ij}(a)e_{jk}(b)e_{ij}(a)^{-1}e_{jk}(b)^{-1}=e_{ik}(ab)\quad i,j,k\:\mbox{distinct}\]
one can split every element in $E(\mathbb{C})$ into a (group) commutator
of two other elements. 

Given a grope $\mathcal{G}$ representing via $\pi_{1}(\mathcal{G})=\pi$
the (superperfect) group $\pi$. Now we define the $C^{*}-$algebra
$C^{*}(\mathcal{G},\pi$) associated to the grope $\mathcal{G}$ with
group $\pi$. The basic elements of this algebra are smooth half-densities
with compact supports on $\mathcal{G}$, $f\in C_{c}^{\infty}(\mathcal{G},\Omega^{1/2})$,
where $\Omega_{\gamma}^{1/2}$ for $\gamma\in\pi$ is the one-dimensional
complex vector space of maps from the exterior power $\Lambda^{2}L$
, of the union of levels $L$ representing $\gamma$ to $\mathbb{C}$
such that \[
\rho(\lambda\nu)=|\lambda|^{1/2}\rho(\nu)\qquad\forall\nu\in\Lambda^{2}L,\lambda\in\mathbb{R}\:.\]
For $f,g\in C_{c}^{\infty}(\mathcal{G},\Omega^{1/2})$, the convolution
product $f*g$ is given by the equality\[
(f*g)(\gamma)=\intop_{[\gamma_{1},\gamma_{2}]=\gamma}f(\gamma_{1})g(\gamma_{2})\]
Then we define via $f^{*}(\gamma)=\overline{f(\gamma^{-1})}$ a $*$operation
making $C_{c}^{\infty}(\mathcal{G},\Omega^{1/2})$ into a $*$algebra.
For each capped torus $T$ in some level of the grope $\mathcal{G}$
one has a natural representation of $C_{c}^{\infty}(\mathcal{G},\Omega^{1/2})$
on the $L^{2}$ space over $T$. Then one defines the representation\[
(\pi_{x}(f)\xi)(\gamma)=\intop_{[\gamma_{1},\gamma_{2}]=\gamma}f(\gamma_{1})\xi(\gamma_{2})\qquad\forall\xi\in L^{2}(T).\]
The completion of $C_{c}^{\infty}(\mathcal{G},\Omega^{1/2})$ with
respect to the norm \[
||f||=\sup_{x\in M}||\pi_{x}(f)||\]
makes it into a $C^{*}$algebra $C_{c}^{\infty}(\mathcal{G},\pi$).
Via the representation $U:\pi\to E(\mathbb{C})$, we get a homomorphism
into the usual convolution algebra $C^{*}(E(\mathbb{C}))$ of the
group $E(\mathbb{C})$ used later to construct the action of the quantum
D-brane. Finally we are able to define the $C^{*}-$algebra associated
to the wild embedding: \begin{definition} Let $j:K\to S^{n}$ be
a wild embedding with $\pi=\pi_{1}(S^{n}\setminus j(K))$ as fundamental
group of the complement $M(K,j)=S^{n}\setminus j(K)$. The $C^{*}-$algebra
$C_{c}^{\infty}(K,j)$ associated to the wild embedding is defined
to be $C_{c}^{\infty}(K,j)=C_{c}^{\infty}(\mathcal{G},\pi)$ the $C^{*}-$algebra
of the grope $\mathcal{G}$ with group $\pi$. \end{definition}

\subsection{Isotopy classes of wild embeddings and KK theory}

In section \ref{sub:Wild-and-tame-embed} we introduce the notion
of isotopy classes for embeddings. Given two embeddings $f,g:N\to M$
with a special map $F:M\times[0,1]\to M\times[0,1]$ as deformation
of $f$ into $g$, then both embeddings are isotopic to each other.
The definition is independent of the tameness or wilderness for the
embedding. Now we specialize to our case of wild embeddings $f,g:K\to S^{n}$
with complements $M(K,f)$ and $M(K,g)$. The map $F:S^{n}\times[0,1]\to S^{n}\times[0,1]$
induces a homotopy of the complements $M(K,f)\simeq M(K,g)$ giving
an isomorphism of the fundamental groups $\pi_{1}(M(K,g))=\pi_{1}(M(K,f))$.
Thus, the isotopy class of the wild embedding $f$ is completely determined
by the $M(K,f)$ up to homotopy. Using Connes work on operator algebras
of foliation, our construction of the $C^{*}-$algebra for a wild
embedding is functorial, i.e. an isotopy of the embeddings induces
an isomorphism between the corresponding $C^{*}-$algebras. Given
two non-isotopic, wild embeddings then we have a homomorphism between
the $C^{*}-$algebras only. But every homomorphism (which is not a
isomorphism) between $C^{*}-$algebras $A,B$ gives an element of
$KK(A,B)$ and vice verse. Thus,

\begin{theorem}

Let $j:K\to S^{n}$ be a wild embedding with $\pi=\pi_{1}(S^{n}\setminus j(K))$
as fundamental group of the complement $M(K,j)=S^{n}\setminus j(K)$
and $C^{*}-$algebra $C_{c}^{\infty}(K,j)$. Given another wild embedding
$i$ with $C^{*}-$algebra $C_{c}^{\infty}(K,i)$. The elements of
$KK(C_{c}^{\infty}(K,j),C_{c}^{\infty}(K,i))$ are the isotopy classes
of the wild embedding $j$ relative to $i$. 

\end{theorem}

\subsection{Wild embeddings are quantum D-branes}

Given a wild embedding $f:K\to S^{n}$ with $C^{*}-$algebra $C^{*}(K,f)$
and group $\pi=\pi_{1}(S^{n}\setminus f(K))$. In this section we
will derive an action for this embedding to get back the D-brane action
in the classical limit. The starting point is our remark above that
the group $\pi$ can be geometrically constructed by using a grope
$\mathcal{G}$ with $\pi=\pi_{1}(\mathcal{G})$. This grope was used
to construct a codimension-1 foliation on the 3-sphere classified
by the Godbillon-Vey invariant. This class can be seen as element
of $H^{3}(BG,\mathbb{R})$ with the holonomy groupoid $G$ of the
foliation. The strong relation between the grope $\mathcal{G}$ and
the foliation gives an isomorphism for the $C^{*}-$algebra which
can be easily verified by using the definitions of both algebras.
As shown by Connes \cite{Connes1984,Connes94}, the Godbillon-Vey
class $GV$ can be expressed as cyclic cohomology class (the so-called
flow of weights)\[
GV_{HC}\in HC^{2}(C_{c}^{\infty}(G))\simeq HC^{2}(C_{c}^{\infty}(\mathcal{G},\pi))\]
of the $C^{*}-$algebra for the foliation isomorphic to the $C^{*}-$algebra
for the grope $\mathcal{G}$. Then we define an expression\[
S=Tr_{\omega}\left(GV_{HC}\right)\]
uniquely associated to the wild embedding ($Tr_{\omega}$ is the Dixmier
trace). $S$ is the action of the embedding. Because of the invariance
for the class $GV_{HC}$, the variation of $S$ vanishes if the map
$f$ is a wild embedding. But this expression is not satisfactory
and cannot be used to get the classical limit. For that purpose we
consider the representation of the group $\pi$ into the group $E(\mathbb{C})$
of elementary matrices. As mentioned above, $\pi$ is countable generated
and the generators can be arranged in the embeddings space. Then we
obtain matrix-valued functions $X^{\mu}\in C_{c}^{\infty}(E(\mathbb{C}))$
as the image of the generators of $\pi$ w.r.t. the representation
$\pi\to E(\mathbb{C})$ labeled by the dimension $\mu=1,\ldots,n$
of the embedding space $S^{n}$. Via the representation $\iota:\pi\to E(\mathbb{C})$,
we obtain a cyclic cocycle in $HC^{2}(C_{c}^{\infty}(E(\mathbb{C}))$
generated by a suitable Fredholm operator $F$. Here we use the standard
choice $F=D|D|^{-1}$ with the Dirac operator $D$ acting on functions
$C_{c}^{\infty}(E(\mathbb{C}))$. Then the cocycle in $HC^{2}(C_{c}^{\infty}(E(\mathbb{C}))$
can be expressed by\[
\iota_{*}GV_{HC}=\eta_{\mu\nu}[F,X^{\mu}][F,X^{\nu}]\]
using a metric $\eta_{\mu\nu}$in $S^{n}$ via the pull-back using
the representation $\iota:\pi\to E(\mathbb{C})$. Finally we obtain
the action\begin{equation}
S=Tr_{\omega}([F,X^{\mu}][F,X_{\mu}])=Tr_{\omega}([D,X^{\mu}][D,X_{\mu}]|D|^{-2})\label{eq:quantum-D-brane-action}\end{equation}
which can be evaluated by using the heat-kernel of the Dirac operator.
For the classical limit, we take a tame embedding $f:K\to S^{n}$
of a $p-$dimensional complex $K$. Then the group $\pi$ simplifies
to a finite group or is trivial. The Dirac operator $D$ on $K$ acts
on usual square-integrable functions and the action simplifies to\[
S=\intop_{K}\left(\eta_{\mu\nu}\partial^{\alpha}X^{\mu}\partial_{\alpha}X^{\nu}+\frac{1}{3}R+\ldots\right)dvol(K)\]
for the main contributions where $R$ is the scalar curvature of $K$
(for $p>2$). It is known that this action agrees with the usual Born-Infeld
action for $p-$branes ($p>2)$ if $R>0$. Thus we obtain a description
of the quantum D-brane action by using wild embeddings for the description
of a quantum D-brane. We will further investigate this point in a
forthcoming paper.

\section{Conclusion}

In this paper we present a lot of results to support our main conjecture:\\
\emph{The exotic small $\mathbb{R}^{4}$ lies at the heart of quantum
gravity. Especially it is a quantized object.}\\
Here we are mainly concentrated on the various relation to branes
in superstring theory as a possible candidate of quantum gravity.
We found the amazing connections between 4-exotics and NS and D-branes
in various string backgrounds. We also studied the case of quantum
D-branes using $C^{*}-$algebras. All the results can be simply summarized
by:\\
\emph{The exotic small $\mathbb{R}^{4}$ as described by codimension-1
foliations on the 3-sphere is the germ of wide range of effects on
D-branes. A quantum Dp-brane is given by a wild embedding of a $p-$dimensional
complex into a $n-$dimensional space described by a two-dimensional
complex, a grope. }\\
Further evidences supported this statement as well the relation with
supersymmetry and realistic QFT will be presented in a separate paper. 

But as known from our previous work, the grope is the main structure
to get the relation between the exotic small $\mathbb{R}^{4}$ and
the codimension-1 foliation on the 3-sphere. The description of the
wild embedding is rather independent of the dimensions ($n>6$, $p>2$).
That is the reason why the exotic small $\mathbb{R}^{4}$ appeared
in so different situations above!

\section*{Acknowledgment}

T.A. wants to thank C.H. Brans and H. Ros\'e for numerous discussions
over the years about the relation of exotic smoothness to physics.
J.K. benefited much from the explanations given to him by Robert Gompf
regarding 4-smoothness several years ago, and discussions with Jan
S{\l}adkowski.

\appendix
\addtocontents{toc}{\protect\setcounter{tocdepth}{0}}
\section{Remarks on K-matrix string theory and BSFT \label{sec:Elements-of-K-matrix}}

D-branes are solitonic objects in string theory similar to instantons
in gauge theory. Both can be seen as generically nonperturbative objects
and one can consider dual theories on backgrounds of these. To investigate
less supersymmetric phases of string theory one can introduce non-BPS
branes. In fact in flat space background of type II string theory
there are two kinds of D-branes systems without supersymmetry. They
are brane-anti-brane systems and non-BPS D-branes. From the point
of view of open string theory on such systems of branes the spectra
of open strings contain tachyon's modes. In the closed string spectra
tachyons do not appear. The appearance of tachyons is the indication
that the theory (without supersymmetry) is in its unstable phase and
will decay into the stable phase where some supersymmetry is restored.
The conjecture by Ashoke Sen states that the unstable D-branes decay
to the closed backgrounds without any brane or to stable D-branes,
without tachyons. This decay process is called tachyon condensation,
and is a dynamical phenomena which can not be described in perturbative
open string theory since we relate two different backgrounds. Similarly
as in various dualities in string theory where different backgrounds
are related, also here we deal rather with important non-perturbative
aspects of the theory. The Ashok conjecture one can check by techniques
of noncommutative field theory, K-theory and field string theory.
This last includes boundary field string theory (BFST) as introduced
by Witten where one can calculate exact expressions for the tachyon
potential and effective actions for the systems of unstable D-branes
(including DBI-like and WZ ones).

In fact any kind of D-brane in type II superstring theory can be obtained
as a soliton in the gauge theory defined on higher dimensional unstable
D-brane systems. The gauge theory is equipped with a tachyon field.
From this descent relation, from unstable D-branes to the lower dimensional
solitons - D-branes, follows, in particular, the classification of
the RR charges of D-branes in terms of K-theory. There is, however,
yet another, ascent way to D-branes: these are realized as bound states
of lower dimensional unstable D-brane systems. In a particular case
of string theory described by Matrix theory and lowest possible dimension
of unstable D-branes, namely D-instantons (D-particles), the construction
of type II D-branes as bound states of D-instantons is possible and
this is what one calls K-matrix theory. The reason is particularly
close and natural connection of K-homology classes (and spectral triples
from noncommutative geometry) with world-volumes of D-branes as we
discussed in this paper. Moreover, the ascent and descent relations
together give rise to the same D-branes hence the suitable K-theory
describing D-branes should be enhanced to the KK-theory of Kasparov
in the context of $C^{\star}$ algebra bi-modules. The tachyon condensation
is crucial for both, ascent and descent relations leading to D-branes
and for the D-instantons case in K-matrix string theory. 

Following \cite{AsakawaSugimotoTerasima2003} let us see how tachyons
and other fields on unstable brane systems emerge from BSFT description.
The boundary states ascribed to a D-brane are linear combinations
of the states $|Bp;\pm>$, which are represented as the formal integration
out of longitudinal fields along the world-volume, i.e. 

\[
|Bp;\pm>=\int[x^{\alpha}][d\psi^{\alpha}]|X^{\alpha},x^{i}=0>|\psi^{\alpha},\psi^{i}=0;\pm>\]
where $\alpha$ are longitudinal and $i$ transversal directions to
the $Dp$-brane. Turning on the fields on the world-volume is via
the boundary interactions and modifies the boundary states as

$|Bp;\pm>_{b}=\int[x^{\alpha}][d\psi^{\alpha}]e^{-S_{b}(x,\psi)}|X^{\alpha},x^{i}=0>|\psi^{\alpha},\psi^{i}=0;\pm>$
where $S_{b}$ is the boundary action of boundary interactions. This
can be seen in the case of gauge fields on the brane as supersymmetric
generalization of operators represented by Wilson loops

$e^{-S_{b}(X,\Psi_{\pm})}={\rm TrP}\exp\left\{ -\int d\sigma\left(A_{\alpha}(X)\dot{X}\right)-\frac{1}{2}F_{\alpha\beta}(X)\Psi_{\pm}^{\alpha}\Psi_{\pm}^{\beta}\right\} $.
Rewriting this in terms of superfields $\hat{X}(\sigma,\theta)=X^{\mu}(\sigma)+i\theta\Psi_{\pm}^{\mu}(\sigma)$,
$\hat{x}^{\mu}(\sigma,\theta)=x^{\mu}(\sigma)+i\theta\sigma_{\pm}^{\mu}(\sigma)$
and the covariant superderivative, $D=\partial_{\theta}+\theta\partial_{\sigma}$,
where $(\sigma,\theta)$ are super coordinates on the boundary, we
get

\[
e^{-S_{b}(X,\Psi_{\pm})}={\rm Tr\tilde{P}}\exp\left(-\int d\sigma d\theta\left(A_{\alpha}(\hat{X})\right)D\hat{X}^{\alpha}\right)\]
 where $\tilde{P}$ is the supersymmetric path-ordered product. 

Now the boundary action can be given for the case of unstable Dp-branes
(IIA type) or the system of non BPS branes - antibranes (type IIB).
We introduce the matrix 

\[
\hat{M}=\left(\begin{array}{cc}
-A_{\alpha}(\hat{X})D\hat{X}^{\alpha}-i\Phi^{i}(\hat{X})\hat{P}_{i} & T(\hat{X})\\
T(\hat{X})^{\dagger} & -A'_{\alpha}(\hat{X})D\hat{X}^{\alpha}-i\Phi'^{i}(\hat{X})\hat{P}_{i}\end{array}\right)\]
here $\hat{P}_{i}=\theta P_{i}(\sigma)+i\Pi_{i\pm}(\sigma)$ and $P_{i}$,
$\Pi_{i\pm}$ are conjugate momenta of $X_{i}$, $\Psi_{i\pm}$ respectively.
$A_{\alpha}$, $A'_{\alpha}$, $\Phi^{i}$, $\Phi'^{i}$, $T$ are
gauge fields, scalar fields and tachyon on the brane and anti brane
which in the case of non-BPS Dp-brane are not independent, i.e. $A_{\alpha}=A'_{\alpha}$
and $\Phi^{i}=\Phi'^{i}$. Matrix $\hat{M}$ is rewritten in terms
of Pauli matrices $\sigma_{1}$, $\sigma_{2}$ and in terms of redefined
fields: $A_{\alpha}^{\pm}=\frac{1}{2}(A_{\alpha}\pm A'_{\alpha})$,
$\Phi_{\alpha}^{\pm}=\frac{1}{2}(\Phi_{\alpha}\pm\Phi'_{\alpha})$,
$T^{\pm}=\frac{1}{2}(T\pm T^{\dagger})$ as:

\[
\hat{M}=-(A_{\alpha}^{+}D\hat{X}^{\alpha}+i\Phi_{+}^{i}P_{i})\otimes1_{2}-(A_{-}^{+}D\hat{X}^{\alpha}+i\Phi_{-}^{i}P_{i})\otimes\sigma_{2}\sigma_{1}+T^{+}\sigma_{1}+T^{-}\sigma_{1}\:.\]

Introducing additional real fermionic superfields 

\begin{equation}
\hat{\Gamma}^{I}(\sigma,\theta)=\eta^{I}(\sigma)+\theta F^{I}(\sigma)\label{eq:superfieldGamma}\end{equation}
 $I=1,2$, we obtain, now for the case of $N$ non-BPS pairs of Dp
- anti-Dp branes, the boundary action:

\[
e^{-S_{b}}=\int[d\hat{\Gamma^{1}}][d\hat{\Gamma^{2}}]{\rm Tr}\tilde{P}\exp\{\int d\sigma d\theta(\frac{1}{4}\hat{\Gamma}^{1}D\hat{\Gamma}^{1}+\frac{1}{4}\hat{\Gamma}^{2}D\hat{\Gamma}^{2}+(A_{\alpha}^{+}D\hat{X}^{\alpha}+i\Phi_{+}^{i}P_{i})+\]

$+(A_{-}^{+}D\hat{X}^{\alpha}+i\Phi_{-}^{i}P_{i})i\hat{\Gamma}^{2}\hat{\Gamma}^{1}+T^{+}\hat{\Gamma}^{1}+T^{-}\hat{\Gamma}^{2}){\rm \}}\:.$

Next one expands $\hat{M}$ in terms of $SO(2m)$ gamma matrices,
namely the antisymmetrizied product $\Gamma^{I_{1}...I_{k}}=\Gamma^{[I_{1}}...\Gamma^{I_{k}]}$ 

\[
\hat{M}=\sum_{k=0}^{2m}\hat{M}^{I_{1...I_{k}}}\otimes\Gamma^{I_{1}...I_{k}}\:.\]

The form of the boundary action follows:

\[
e^{-S_{b}}=\int[d\hat{\Gamma^{I}}]{\rm Tr}\tilde{P}\exp\left\{ \int d\sigma d\theta\left(\frac{1}{4}\hat{\Gamma}^{I}D\hat{\Gamma}^{I}+\sum_{k=0}^{2m}\hat{M}^{I_{1...I_{k}}}\otimes\hat{\Gamma}{}^{I_{1}...I_{k}}\right)\right\} \:.\]

The $\theta$ integration can be performed which after integrating
out the $F$ fields in (\ref{eq:superfieldGamma}), gives

\[
e^{-S_{b}}=\int[d\eta^{I}]{\rm Tr}\tilde{P}\exp\left\{ \int d\sigma\left(\frac{1}{4}\eta^{I}\dot{\eta}{}^{I}+\sum_{k=0}^{2m}M^{I_{1...I_{k}}}\otimes\Gamma{}^{I_{1}...I_{k}}\right)\right\} \:.\]

This last path integral expression is equivalent to the operator one,
after integrating on $d\eta$ and in the superfields formalism the
action reads:

\begin{equation}
e^{-S_{b}}=\begin{array}{c}
{\rm Tr\tilde{P}}e^{\int d\sigma d\theta\hat{M}(\sigma)},\:{\rm for}\:{\rm Dp-anti-Dp-branes}\\
\frac{1}{\sqrt{2}}{\rm Tr\tilde{P}}e^{\int d\sigma d\theta\hat{M}(\sigma)},\:{\rm for}\:{\rm non-stable}\:{\rm Dp-branes}\end{array}\:.\label{eq:BoundaryAction}\end{equation}

which in terms of fields is (in the NS sector again):

\begin{equation}
e^{-S_{b}}=\begin{array}{c}
{\rm TrP}e^{\int d\sigma M(\sigma)},\:{\rm for}\:{\rm Dp-anti-Dp-branes}\\
\frac{1}{\sqrt{2}}{\rm TrP}e^{\int d\sigma M(\sigma)},\:{\rm for}\:{\rm non-stable}\:{\rm Dp-branes}\end{array}\:.\label{eq:BoundaryActionII}\end{equation}

Turning to the D-instantons case let us consider $N$ non BPS D-instantons
which are the lowest dimensional D-branes in type IIA theory. The
gauge theory on such systems is $U(N)$ gauge theory. No gauge field
$A$ are present but bosons of the theory consist of 10 scalar fields
$\Phi^{\mu}$, $\mu=0,\,1,...,9$ and tachyon $T$ which are from
the adjoint representation of $U(N)$. One takes limit $N\to\infty$
since creation or annihilation of arbitrary many of non-BPS D-instantons
can be considered then. Thus the infinite Hermitian matrices $\Phi^{\mu}$,
$T$ represent the linear operators acting on the separable Hilbert
space ${\cal H}$. The BSFT action for such system for the NSNS sector,
reads

\begin{equation}
S(\Phi^{\mu},T)=\frac{2\pi}{g_{s}}<0|e^{-S_{b}(\Phi^{\mu},T)}|B(-1);+>_{NS}\end{equation}
Here $S_{b}(\Phi^{\mu},T)$ is the action governing the interaction
of the boundary states from closed strings Hilbert space ${\cal H}$
as before. The matrix $\hat{M}$ in (\ref{eq:BoundaryAction}) is
now given as

\[
\left(\begin{array}{cc}
-i\Phi^{\mu}\hat{P}_{\mu} & T\\
T & -i\Phi^{\mu}\hat{P}_{\mu}\end{array}\right)\]
or $M$ in (\ref{eq:BoundaryActionII}), as

\[
\left(\begin{array}{cc}
-i\Phi^{\mu}P_{\mu}-T^{2}-\frac{1}{2}[\Phi^{\mu},\Phi^{\nu}]\Pi_{\mu}\Pi_{\nu} & -[\Phi^{\mu},T]\Pi_{\mu}\\
-[\Phi^{\mu},T]\Pi_{\mu} & -i\Phi^{\mu}P_{\mu}-T^{2}-\frac{1}{2}[\Phi^{\mu},\Phi^{\nu}]\Pi_{\mu}\Pi_{\nu}\end{array}\right)\:.\]

The very important thing is that any D-brane configuration of IIA
string theory can be constructed in the above K-matrix string theory
from D-instantons. Thus, solution representing Dp-brane is the following
configuration:

\begin{equation}
T=u\sum_{\alpha=0}^{p}\tilde{p}_{\alpha}\otimes\gamma^{\alpha},\:\Phi^{\alpha}=\tilde{x}^{\alpha}\otimes1,\:\alpha=0,...,p,\:\Phi^{i}=0,\: i=p+1,...,9\end{equation}
which becomes an exact solution in the limit $u\to\infty$, and $\tilde{p}$,
$\tilde{x}$ are operators acting on ${\cal H}$. The eigenvalues
of $\Phi^{\mu}$ represent the position of non-BPS D-instantons and
Dp-brane extends over $0,...,p$ directions. In the case of $N$ Dp-branes
one has:

\[
T=u\sum_{\alpha=0}^{p}\tilde{p}_{\alpha}\otimes1_{N}\otimes\gamma^{\alpha},\:\Phi^{\alpha}=\tilde{x}^{\alpha}\otimes1,\:\alpha=0,...,p,\:\Phi^{i}=0\]
the Hilbert space for $T$ and $\Phi^{\mu}$ is ${\cal H}\otimes\mathbb{C}^{N}\otimes S$
and $S$ is spinor space representing $\gamma^{\alpha}$. The gauge
fields corresponding to the $U(N)$ symmetry, which is now a field
on Dp-brane world-volume, reappear and modify the tachyon operator
as 

\begin{equation}
T=u\sum_{\alpha=0}^{p}(\tilde{p}_{\alpha}\otimes1_{N}-iA_{\alpha}(\tilde{x}))\otimes\gamma^{\alpha}\:.\label{eq:tachyonDirac1}\end{equation}
We see the close connection of tachyon and Dirac operators.

\section{Tachyons and D brane charges \label{sec:Spectral-triples-and}}

The Chern-Simons term in BSFT is obtained by considering the state
$<C|$ in closed string theories corresponding to the RR field $C$,
which couples to the boundary state $|Bp;+>$ representing the Dp-brane,
i.e.

\[
S_{CS}(C,T,A_{\alpha},\Phi^{i})=<C|e^{-S_{b}}|Bp;+>_{RR}\]

Again, following \cite{AsakawaSugimotoTerasima2003}, the relation
between non-stable D(-1)-branes (D-instantons) in K-string theory
with Dp-branes in type II seen from BSFT formalism, allows to state
the equality:

\begin{equation}
S_{CS}^{D(-1)}(C,T,\Phi^{\mu})=S_{CS}^{Dp}(C,A_{\mu},t,\phi^{i})\label{eq:CS1}\end{equation}
where $,A_{\mu},t,\phi^{i}$ are rather fields on the world-volume
of the Dp-brane than operators on Hilbert space as on the lhs. of
(\ref{eq:CS1}). This equality of couplings to RR fields, i.e.Chern-Simons
terms, one from operator side and the second as in gauge theory, gives
rise to a deep relation which illustrates the Atiyah-Singer index
theorem. Let us note that when Dp-brane is BPS the tachyon field $t$
is absent and when we additionally nullify the scalar fields $\phi^{i}$,
the CS term becomes usual one as in gauge theory, i.e. $S_{CS}^{Dp}(C,A_{\mu})=\mu_{p}\int_{Dp}C\wedge\mbox{{\rm tr}}e^{2\pi F}$. 

The Chern-Simons term for the non-BPS D-instantons is given as (\cite{AsakawaSugimotoTerasima2003},
p. 27):

\begin{equation}
S_{CS}^{D(-1)}=<\psi_{2}^{\mu}=0|d^{10}kC(k^{\mu},\psi_{1}^{\mu}){\rm Str}(e^{-ik_{\mu}\Phi^{\mu}+2\pi{\cal F}})|\psi_{1}^{\mu}=0>\end{equation}

where Str is the supersymmetrized trace and

\[
{\cal F}=-T^{2}+\frac{1}{8\pi^{2}}[\Phi^{\mu},\Phi^{\nu}]\psi_{2}^{\mu}\psi_{1}^{\mu}-\frac{i}{2\pi}[\Phi^{\mu},T]\psi_{2}^{\mu}\sigma_{1}\:.\]

The similar CS term for the pairs of D-instantons - anti-D-instantons
from IIB type K-matrix string theory, reads (\cite{AsakawaSugimotoTerasima2003},
p. 28-29). \begin{equation}
S_{CS}^{D(-1)}=<\psi_{2}^{\mu}=0|d^{10}kC(k^{\mu},\psi_{1}^{\mu}){\rm Str}(e^{-ik_{\mu}\tilde{\Phi}^{\mu}+2\pi\tilde{{\cal F}}})|\psi_{1}^{\mu}=0>\end{equation}
here \begin{equation}
\tilde{{\cal F}}=\left(\begin{array}{cc}
-TT^{\dagger}+\frac{1}{8\pi^{2}}[\Phi^{\mu},\Phi^{\nu}]\psi_{2}^{\mu}\psi_{2}^{\mu} & -\frac{i}{2\pi}(\Phi^{\mu}T-T\Phi'^{\mu})\psi_{2}^{\mu}\\
-\frac{i}{2\pi}(\Phi'^{\mu}T^{\dagger}-T^{\dagger}\Phi{}^{\mu})\psi_{2}^{\mu} & -T^{\dagger}T+\frac{1}{8\pi^{2}}[\Phi^{\mu},\Phi^{\nu}]\psi_{2}^{\mu}\psi_{2}^{\mu}\end{array}\right)\end{equation}

Supposing $k_{\mu}=0$ and all components but $C_{0}$ vanish, the
scalar fields $\Phi^{\mu}$, $\Phi'^{\mu}$ vanish too. Thus we have:

\[
S_{CS}^{D(-1)}=C_{0}{\rm Str}\exp\left(-2\pi\left(\begin{array}{cc}
0 & T\\
T^{\dagger} & 0\end{array}\right)^{2}\right)=C_{0}{\rm Ind}\left(\left(\begin{array}{cc}
0 & T\\
T^{\dagger} & 0\end{array}\right)\right)\]
which means $S_{CS}^{D(-1)}=C_{0}(\dim\ker TT^{\dagger}-\dim\ker T^{\dagger}T)$,
i.e. the index of the tachyon operator is interpreted as D-instanton
charge. 

When BPS Dp-brane is present the tachyon operator, following (\ref{eq:tachyonDirac1}),
is

\[
u\sum_{\alpha=0}^{p}(\tilde{p}^{\alpha}-iA_{\alpha}(\tilde{x}))\Gamma^{\alpha}=-iuD\]
where $D$ is the Dirac operator and $\Gamma^{\alpha}$ are $p+1$
$SO(p+1)$ gamma matrices for odd $p$. We see that the D(-1) charge
in the presence of Dp-brane is just the index of the Dirac operator
defined on the world-volume of Dp-brane.

From the equality (\ref{eq:CS1}) of CS terms and the coupling $S_{CS}^{Dp}(C,A_{\mu})=\mu_{p}\int_{Dp}C\wedge\mbox{{\rm tr}}e^{2\pi F}$,
we have

\[
{\rm index}(-iD)=\int_{Dp}{\rm tr}e^{F/2\pi}\]
which is nothing but the variant of the Atiyah-Singer index theorem.
Thus (\cite{AsakawaSugimotoTerasima2003}, p. 30), \emph{the Dirac
operator is the tachyon operator representing the Dp-brane in the
system of D-instanton - anti-D-instanton}. \emph{The index of the
Dirac operator is the charge of D-instantons. The D-instanton can
be equivalently described as the ordinary instantons configuration
in the gauge theory defined on the world-volume of Dp-brane. In that
case the Chern number of the gauge bundle on the world-volume is the
instanton number. }

The above observation opens the possibility to interpret D-branes
as spectral triples also in more general non-commutative situations.
The appearance of the Atiyah-Singer theorem in the context of charges
of Dp-branes can be understood also quite generally via relation of
the charges with K-theory classes and the duality of K-theory with
K-homology \cite{Szabo1999}. Namely let ${\cal T}$ be a bounded
linear operator acting on a separable Hilbert space ${\cal H}$ with
kernel and cokernel being finite dimensional vector spaces, i.e. a
Fredholm operator. The index of it is 

\begin{equation}
{\rm index}\,{\cal T}=\dim\ker{\cal T}-\dim{\rm coker}\,{\cal T}\label{eq:index1}\end{equation}

Let ${\cal F}$ be the space of Fredholm operators on ${\cal H}$
with the norm (of operators) topology. From (\ref{eq:index1}) a continuous
map ${\rm index:}\:{\cal F}\to\mathbb{Z}$ is defined which is in
fact a bijection between connected components of ${\cal F}$ and $\mathbb{Z}$.
For given $X$ a compact topological space the set $[X,{\cal F}]$
of homotopy classes of maps $X\to{\cal F}$ with its monoid structure
is isomorphic to K-theory classes:

\begin{equation}
[X,{\cal F}]\simeq K(X)\label{eq:K-theory-Fredholm}\end{equation}
This is basically since the kernel and cokernel of the continuous
family of the Fredholm operators on $X$ acting on ${\cal H}$, are
vector bundles over $X$. Thus the K-theory class of the above pair
of vector bundles derived from the above family of Fredholm operators,
is determined. Let denote it as ${\rm Index}\,{\cal T}:=\left[(\ker{\cal T},{\rm coker}\,{\cal T})\right]\in K(X)$
\cite{Szabo1999}.

For a point $pt$ we have $K(pt)=\mathbb{Z}$ and the Index defined
via K-theory of the family of Fredholm operators (\ref{eq:K-theory-Fredholm})
becomes exactly the index of the Fredholm operator (\ref{eq:index1}).
The K-homology group $K_{0}(X)$, which is the set of homotopy classes
of Fredholm operators $\left\{ [{\cal P}]\right\} $, is dual to the
K-theory and the duality is given by the pairing 

\begin{equation}
\left([E],[{\cal P}]\right)\to{\rm index}(P_{E})\in\mathbb{Z}\:.\label{eq:pairing1}\end{equation}
Here $E$ is the vector bundle $E\to X$ representing the class $[E]\in K(X)$
and ${\cal P}:\Gamma(X,E)\to\Gamma(X,E)$ is the Fredholm operator
defined on the Hilbert space ${\cal H}=L^{2}(\Gamma(X,E))$. Let us
observe now that Dirac operator defined on a spin manifold with vector
bundle fits to this schema since it is Fredholm on a suitable Hilbert
space \cite{Szabo1999}. Namely $iD:\Gamma(X,S_{E}^{+})\to\Gamma(X,S_{E}^{-})$
where $E\to X$ is a real spin bundle and $S_{E}^{\pm}$ are twisted
by $E$ spinor bundles. Now the Chern character is defined for K-theory
${\rm ch}:K(X)\to H^{\star}(X,\mathbb{Q})$, taking it to the cohomology
classes. In the case of smooth manifold and a smooth vector bundle
$E\to X$ with a Hermitian connection with the curvature $\nabla_{E}^{2}=F_{E}$,
the Chern character reads 

\[
{\rm ch}(E)={\rm tr}\, e^{F_{E}/2\pi i}\,.\]
We can determine the analytic index of $iD$ via K-theory as before.
Then using the Chern character (and Gysin map) one translates the
analytic index to the topological one, via cohomology classes, which
is just the Atiyah-Singer index theorem \cite{Szabo1999,Valentino2008}

\[
{\rm index}\,(-iD)=\int_{X}{\rm ch}(E)\wedge\hat{A}(TX)\,.\]

In terms of K-theory there is a natural bi-linear pairing given by
the index of the twisted Dirac operator associated to the ${\rm spin}^{c}$
structure on $X$ ($E$, $F$ being vector bundles representing the
K-theory classes):

\begin{equation}
(E,F)_{K}=index(D_{E\otimes F})\,.\label{eq:K-pairing}\end{equation}

The Chern character gives rise to the ring isomorphism

\[
{\rm ch}:K(X)\otimes\mathbb{Q}\simeq H(X,\mathbb{Q})\]
which should be modified as ${\rm ch}\to\sqrt{{\rm Td}(X)}\smile{\rm ch}$
in order to preserve (\ref{eq:K-pairing}). Here Td is the Todd (reversible)
class of the tangent bundle of $X$. The above modification is again
nothing but the Atiyah-Singer index theorem \cite{Szabo2008a}:

\[
index(D_{E\otimes F})=C\int_{X}{\rm ch}(E\otimes F)\smile{\rm Td}(X)\,.\]

The above duality and generalization of index theorem can be extended
over noncommutative spaces given by $C^{\star}$-algebras leading
to the suitable formula for the charge of noncommutative branes \cite{Szabo2008a,Szabo2008b,Szabo2008c}.

\addcontentsline{toc}{section}{\refname}

\end{document}